\documentclass[%
 reprint,
 amsmath,amssymb,
 aps,
nofootinbib
]{revtex4-2}

\usepackage{graphicx}% Include figure files
\usepackage{dcolumn}% Align table columns on decimal point
\usepackage{bm}% bold math
\usepackage{array}
\newcommand{\E}{\mathcal{E}}
\newcommand{\calL}{\mathcal{L}}

\newcommand{\vinf}{v_\infty}
\newcommand{\lmo}{{\ell m\omega}}
\newcommand{\lm}{{\ell m}}

\newcommand{\vect}[1]{\ensuremath{\mathbf{#1}}}

\usepackage{orcidlink}

\begin{document}

\preprint{APS/123-QED}

\title{Gravitational radiation from hyperbolic orbits: comparison between self-force, post-Minkowskian, post-Newtonian, and numerical relativity results}% Force line breaks with \\

\author{Niels Warburton\,\orcidlink{0000-0003-0914-8645}}
\affiliation{School of Mathematics and Statistics, University College Dublin, Belfield, Dublin 4, Ireland}

\date{\today}% It is always \today, today,

\begin{abstract}
In this work I use a frequency-domain Regge-Wheeler-Zerilli approach to compute the gravitational wave energy radiated by a compact body moving along a hyperbolic or parabolic geodesic of a Schwarzschild black hole. 
I compare my results with the latest post-Minkowskian (PM) calculations for the radiated energy and find agreement for hyperbolic orbits with large impact parameters and characterized by a velocity at infinity, $\vinf$, as large as $\vinf/c=0.7$.
I also find agreement between my results and the leading-order PM expansion for the radiation absorbed by the black hole.
I make further comparisons with post-Newtonian (PN) theory and show the effectiveness of a simple PN-PM hybrid model.
Finally, I make a first comparison of the radiated energy between self-force and numerical relativity.
\end{abstract}

\maketitle

%\tableofcontents

% \section{Ideas}
% \begin{itemize}
%     \item Comparison with PN and PM for fixed $r_\text{min}$
%     \item Comparison with 5PM for a variety of $v_\infty$ values
%     \item Extract the PN behaviour at 6PM?
%     \item Look at the horizon flux? What is the scaling with b (for fixed $v_\infty$?). Looks like the leading PM horizon flux is calculated here: \cite{Goldberger:2020wbx,Jones:2023ugm}. The horizon flux also enters at 3PM. Can I extract the 4PM coefficient? Need to actually output the horizon flux. Try at $v_\infty/c=0.35$ Looks to work.
% 	\item Do all $(l,m)$ contribute to each PM order? In PN higher-order modes start at higher-order in the PN expansion.
% 	\item compute the radiative scattering angle $\chi^{\rm rad}$. This means outputting the radiated angular momentum. Should output that and the weight coeffs (as they can construct anything in post processing).
% \end{itemize}

\section{Introduction}

Relativistic modelling of compact binaries has seen an explosion of interest in the past decades driven by the needs of gravitational wave (GW) observatories.
Most modelling focuses on bound systems which make up all of the GW sources detected to date \cite{LIGOScientific:2025slb}.
There has been less attention on unbound, scattered binaries as they are not expected to form a significant population of detectable GW sources (though see, e.g., \cite{Berry:2012im,Oliver:2023xan,Henshaw:2025arb}).
Recently, new approaches to modelling scattered binaries by leveraging techniques developed in the particle physics community, along with ways to map the information from unbound to bound systems, had led to a flurry of new work in this area.

There are a variety of techniques to model relativistic binaries.
One can solve the full Einstein field equations on supercomputers in an approach known as Numerical Relativity (NR).
If the binary is asymmetric then the field equations can be perturbatively expanded in the small mass ratio in what is known as the self-force (SF) approach.
In the weak field an alternative perturbative approach, known as the post-Minkowskian (PM) expansion, can be employed whereby one expands in powers of the gravitational constant. 
If in addition the velocity of the binary is also small compared to the speed of light then the post-Newtonian (PN) expansion becomes valid.
All of these approaches have involved decades of theoretical development, and some require the execution of tens of thousands of lines of computer code to produce results.
Given all this complexity it is crucial to cross-check the output from each approach against the others.
The gauge freedom of general relativity can make such cross-checks challenging and so comparisons focus on a handful of gauge invariant quantities.
These gauge invariant quantities also play a key role in calibrating effective models of the binaries, such the effective-one-body (EOB) approach \cite{Buonanno:1998gg, Pompili:2023tna, Riemenschneider:2021ppj}.
There is a rich and fruitful history of these comparisons and calibrations for bound systems \cite{Boyle:2007ft,Hannam:2007wf,Detweiler:2008ft,Blanchet:2009sd,Favata:2010yd,LeTiec:2011bk,Dolan:2013roa,LeTiec:2014oez,Dolan:2014pja,Akcay:2015pza,Antonelli:2019fmq,Antonelli:2019ytb,Borhanian:2019kxt,vandeMeent:2023ols,Leather:2025nhu}.
The primary focus of the present work is to extend cross-validation with SF results to scattered binaries.

In particular, this work is motivated by the recent rapid progress in PM calculations.
Following the seminal work by Damour \cite{Damour:2016gwp, Damour:2017zjx} showing how to map PM results for scattered binaries to bound binaries via the effective-one-body (EOB) approach, a variety of authors have worked to extend PM calculations, often using particle physics methods (see, e.g., \cite{Bjerrum-Bohr:2022blt,Kosower:2022yvp,Buonanno:2022pgc} for reviews).
One recent major result was the calculation of the scattering angle and radiated energy at the fifth PM order and at leading order in the mass ratio expansion \cite{Driesse:2024feo}.
This analytic calculation involved the evaluation of hundreds of Feynman diagrams which was assisted by around 300,000 CPU hours of computational resources.
The resulting formulae contain thousands of terms.
Both their scattering angle and radiated energy formula were validated in the low-velocity limit by comparison with PN, and the scattering angle was validated at higher velocities by comparison with NR simulations.
The present work provides the first validation of PM radiated energy formulae at higher velocities by comparing against SF results.
The agreement I find between PM and SF results builds confidence in both calculations.

I now review the status of each approach for binaries on hyperbolic or parabolic orbits, with an emphasis on the radiated energy which is the focus of the present work.
As I give a more detailed review of the analytic PN and PM calculations in Sec.~\ref{sec:PN-PM-summary} I only briefly review these results here.
PN calculations have been made through 3PN order \cite{Blanchet:1989cu,Bini:2020hmy,Cho:2022pqy}, and PM results are known at all orders in the mass ratio at 3PM \cite{Herrmann:2021lqe} and 4PM \cite{Dlapa:2022lmu,Damgaard:2023ttc} order, and the 5PM results are at known leading order in the mass ratio \cite{Driesse:2024feo}.
The radiation absorbed by the black holes during the scattering event is known at leading order \cite{Goldberger:2020wbx,Jones:2023ugm}.
There is also a range of results made using a double PM-PN expansion \cite{Cho:2021onr,Cho:2022pqy}  and results that also include a third SF expansion \cite{Bini:2024icd, Geralico:2025rof}.
Approaches that map these analytic for scattered binaries results to the bound case include the `boundary-to-bound' (B2B) map \cite{Kalin:2019rwq,Kalin:2019inp,Cho:2021arx,Adamo:2024oxy} and calibration of EOB models \cite{Khalil:2022ylj,Buonanno:2024vkx}.

There have been a variety of studies of scattered binaries in NR, often with a focus on comparing the scattering angle with PM results \cite{Damour:2014afa,Rettegno:2023ghr, Long:2025tvk,Long:2025nmj,Clark:2025kvu, Kogan:2025vml}. 
This includes testing the B2B map \cite{Kankani:2024may}, resummations of PM results \cite{Swain:2024ngs}, and surrogate modelling of the waveforms \cite{Fontbute:2024amb}.

For SF calculations time-domain Teukolsky codes \cite{Sundararajan:2007jg,Sundararajan:2008zm,Field:2020rjr,Harms:2014dqa,Harms:2013ib}  can calculate the gravitational radiation for arbitrary orbital configurations.
This includes hyperbolic orbits and associated comparisons with EOB models \cite{Placidi:2021rkh, Albanesi:2021rby, Albanesi:2022ywx,Faggioli:2024ugn}.
Parabolic orbits have also been studied using Regge-Wheeler-Zerilli (RWZ) perturbation of Schwarzschild black holes in the time domain \cite{Martel:2003jj}.
Frequency-domain (FD) approaches have also been pursued \cite{Oohara:1984ck,1983PhLA...98..407O,Kojima:1984cc}, notably recently by Hopper with RWZ perturbations \cite{Hopper:2017iyq} who also made comparisons with 1PN results \cite{Hopper:2017qus}.
In the FD there have also been studies of related calculations where the two black holes merge \cite{Oohara:1983xip,Kojima:1984cj,Berti:2010ce,Yin:2025kls}.

SF calculations in the conservative sector are significantly more challenging as one needs to both compute and regularize the metric perturbation near the secondary \cite{Pound:2021qin}.
To date all such calculations for hyperbolic orbits have focused on scalar-field toy models.
These include calculations in the time \cite{Barack:2022pde} and frequency domains \cite{Whittall:2023xjp, Whittall:2025dqn}, hybridization of these results \cite{Long:2024ltn}  and comparisons with scalar-field PM results \cite{Barack:2023oqp}.

This work focuses on calculating the radiated energy from a hyperbolic orbit and is structured as follows.
I describe the physical setup of the problem in Sec.~\ref{sec:physical_setup}, and summarize the known analytic results in Sec.~\ref{sec:PN-PM-summary}.
In Sec.~\ref{sec:hybrid} I present a simple hybrid PN-PM model.
Hyperbolic geodesics of Schwarzschild time are described in Sec.~\ref{sec:geodesics}, and the frequency domain method for calculating the associated perturbation is presented in Sec.~\ref{sec:SF_calc}.
In Sec.~\ref{sec:numerical_method} I outline the numerical method, error analysis, and internal validation of the numerical results.
The main results comparing my SF calculation with PM theory are presented in Sec.~\ref{sec:results_comparisons}.
I also take the opportunity here to make comparisons with PN theory, the PN-PM hybrid, and the results of NR simulations.
I provide some concluding remarks and thoughts on future directions in Sec.~\ref{sec:conclusions}, and Appendix \ref{apdx:parabolic} presents a sample of numerical data and results for parabolic orbits.

Throughout this work I use metric signature $(-,+,+,+)$ and I will usually work with geometrized units such that $G=c=1$, unless it adds clarity to restore these factors.
\subsection{Physical setup}\label{sec:physical_setup}

In this work I consider a system where two black holes of mass $m_1 \ge m_2$ scatter off one another.
It will be convenient to consider various combinations of these masses and I define total mass, the small mass ratio, and the symmetric mass ratio, respectively, by
%Standard definitions
\begin{align}
	M &= m_1 + m_2, \\
	\epsilon &= \frac{m_2}{m_1}, \\
	\nu &= \frac{m_1 m_2}{M^2} = \epsilon + \mathcal{O}(\epsilon^2).
\end{align}
In the infinite past the two black holes have \textit{incoming} four-momenta $p^\mu_i = (E_i, \vect{p}_i)$ where $E_i$ and $\vect{p}_i$ are the energy and three-momenta associated with each black hole. 
The magnitude of the four-momentum is given by $p^2_i = - m_i^2$ and I define the total energy as $E = E_1 + E_2$.
In the infinite past the perpendicular distance between the paths the two black holes would take in flat spacetime is the impact parameter, $b$.

% For some of the below see Sec. 2.1 of 1910.03008
When describing the motion of the two bodies there are two key frames to be considered. 
The first is the initial centre of mass (CoM) frame where PN, PM and NR results tend to be computed.
I will denote all quantities in this frame using an asterisk superscript.
The CoM frame is then defined by $\vect{p}^*_2 = - \vect{p}_1^*$.
The four momentum is related to the three-velocity $\vect{v}^*_i$ via $\vect{p}^*_i = \gamma_i^* m_i \vect{v}^*_i$ where $\gamma^*_i = 1/\sqrt{1-(v^*_i)^2}$ is the Lorentz factor with $v^*_i = |\vect{v}^*_i|$.
I define the angular momentum of each body about the CoM via $\vect{L}^*_i = \vect{r}^*_i \times \vect{p}^*_i$, where $\vect{r}^*_i$ is a radial three-vector connecting the body and the CoM.
This gives the magnitude of the angular momentum as $L_i^* \equiv |\vect{L}^*_i| = |\vect{p}_i^*| |\vect{r}^*_i|\sin\theta^*$ where $\theta^*$ is the angle between the two vectors.
The impact parameter between each body and the CoM is defined by $b_i = \lim_{|\vect{r}_i^*| \rightarrow \infty} |\vect{r}^*_i|\sin\theta^*$, and then $b_1 + b_2 = b$.
The magnitude of the total angular momentum is given by $L^* = L_1^* + L_2^* = |\vect{p}_1^*| b$.

Self-force calculations tend to be made in frame where the primary, $m_1$, is stationary.
Boosting to this frame, the initial \textit{relative} velocity and the associated \textit{relative} Lorentz factor between the two black holes is given by
% Agrees with (4.5) of 2010.01641
\begin{equation}\label{eq:v_and_rel_gamma}
	\vinf = \frac{v_1^* + v^*_2}{1 + v^*_1 v^*_2},\qquad \gamma = \frac{1}{\sqrt{1 - \vinf^2}},
\end{equation}
respectively.
Here I use an $\infty$-subscript as this is the velocity of $m_2$ when it is infinitely far from $m_1$.
In this frame the momenta are given by $p^\mu_1 = (m_1,0)$ and $p_2^\mu = (\gamma m_2, \gamma m_2 \vect{v}_\infty)$ which implies $p_1 \cdot p_2 = \gamma m_1 m_2$.
Noting that $E^2 = (p_1^\mu + p_2^\mu)^2$ one can derive the relation
% Agrees with, e.g., Eq. (2b) of 2402.12342
\begin{align}
	\gamma = \frac{E^2 - m_1^2 - m_2^2}{2m_1 m_2}.
\end{align}
%\NW{Note that this $\gamma = u_1 \cdot u_2 = p_1 \cdot p_2 /(m_1 m_2)$?}
From this one can define
% Agrees with, e.g., Eq. (2a) of 2402.12342
\begin{align}
	\Gamma \equiv \frac{E}{M} = \sqrt{1 + 2\nu(\gamma -1)} = 1 + \mathcal{O}(\epsilon).
\end{align}
In the frame where $m_1$ is stationary the angular momentum of $m_2$ about $m_1$ is given by $\vect{L} = \vect{L}_2 = \vect{r}_{12} \times \vect{p}_2$ where $\vect{r}_{12}$ is a vector connecting the two bodies.
The magnitude of the angular momentum $L \equiv |\vect{L}| = |\vect{p}_2| |\vect{r}_{12}|\sin\theta $ where $\theta$ is the angle between the two vectors.
In this frame the impact parameter is given by $b = \lim_{|\vect{r}_{12}| \rightarrow \infty} |\vect{r}_{12}|\sin\theta$.

Using $b = L/|\vect{p}_1|  = L^*/|\vect{p}_1^*|$ and explicit computations of $|\vect{p}_1|$ and $|\vect{p}^*_1|$ one arrives at
% First equation agrees with Eq. (4) of 2511.10196, the second result agrees with Eq. (9) of 2305.09724
\begin{align}\label{eq:b_from_E_L}
	b = E\frac{L^*/(m_1 m_2)}{\sqrt{\gamma^2 - 1}} = \frac{L/m_2}{\sqrt{\gamma^2 -1}}.
\end{align}
In particular this implies that
\begin{equation}
	L^* = \frac{m_1}{E} L = L + \mathcal{O}(\epsilon).
\end{equation}

\section{Summary of analytic results for the gravitational radiation}\label{sec:PN-PM-summary}

%\NW{Should I restore factors of $G$ and $c$ in this section?}
Over the years various PN and PM results for the gravitational radiation from hyperbolic and parabolic orbits have appeared in the literature.
In this section I summarize the status of these results.
For all of these results to be valid the orbital trajectories must remain in the weak field. 
In particular we must have
\begin{align}\label{eq:weakfield}
    \frac{G M}{r c^2} \ll 1.
\end{align}
The PM results will be valid so long as this constraint is satisfied.
For the PN results to be valid the constraint $\vinf/c \ll 1$ must also hold.

% \subsection{Various limits}
%
% From \cite{Barack:2022pde} the weak field limit is given by
% \begin{align}
%     \frac{m_1}{\vinf^2 b} \ll 1
% \end{align}
% This is where the PM expansions are valid.
% The PN limit requires both weak field and $\vinf/c \ll 1$?
% Expanding the PN result in large $e$ (or small angle of deflection) gives the ``bremsstrahlung'' limit.
% Parabolic is $e=1$.

\subsection{Post-Newtonian results}

Through 2PN the radiated energy to infinity takes the form
% Agrees with Eq. (5.7) of Blanchet and Schäfer 1989, and Eq. D2 of arXiv:2007.11239
\begin{align}\label{eq:DeltaE_PN}
	\Delta E^{\rm PN}_\infty(e_r, j) =& \frac{2}{15}\frac{\nu^2}{j^7} \sum_{n=0}^2\frac{1}{j^{2n}}\Bigg[ \mathcal{E}^{\rm nPN}_1\arccos\Bigg(-\frac{1}{e_r}\Bigg)  \nonumber\\
									  &  + \mathcal{E}^{\rm nPN}_2 \sqrt{e_r^2 - 1}\Bigg],
\end{align}
where $j=L^*/M$ and $e_r$ is given through 2PN order in Eq.~(6.20) of \cite{Bini:2020hmy}.
The $\mathcal{E}^{\rm nPN}_k$ (with $k=1$ or $2$) functions appear at the $n$-th PN order, i.e., $\mathcal{O}(c^{-2n})$.
The leading-order Newtonian (0PN) result was computed by Hansen \cite{Hansen:1972jt} (corrected by Turner \cite{1977ApJ...216..610T}).
The 1PN result was computed by Blanchet and Schäfer \cite{Blanchet:1989cu} and the 2PN result was derived by Bini \textit{et al}.~\cite{Bini:2020hmy}; see their Appendix D for the form that matches Eq.~\eqref{eq:DeltaE_PN}.

One often explored limit of these results is the $e_r\rightarrow\infty$, or bremsstrahlung, limit which is equivalent to the large-$j$ limit.
In this limit the radiated energy can be written as
% Agrees with Eq. (C2) of arXiv:2203.10872
% The form below matches Eq. (11.3) of 2107.08896, which also defines p_\infty. Not sure why this differs by a factor of \Gamma from 1911.09130.
% The difference comes from the defintion used in Eq. (3.37) of 2107.08896 where the m1 m2/E is factored out.
\begin{align}\label{eq:DeltaE_PM_PN}
	\Delta E^{\rm PN}_\infty(p_\infty, \hat{\jmath}) =& M \nu^2 \left[\frac{E_3(p_\infty)}{\hat{\jmath}^3} + \frac{E_4(p_\infty)}{\hat{\jmath}^4} + \dots \right],
\end{align}
where $\hat{\jmath} = j^n/p_\infty^{7-n}$ and $p_\infty = \sqrt{\gamma^2 -1}/\Gamma$ is the (specific) linear momentum at infinity. % Defined succinctly in Eq (2.6) of 1911.09130. Also in (C1) of 2203.10872. I should link this to the definitions of momentum that I use above.
The functions $E_n$ arise at the $n\rm PM$ order, i.e., $\mathcal{O}(G^n)$.
At each PM order the function $E_n$ functions can be expanded as
\begin{align}
	E_n = \sum_{k=0}^\infty A_k(\nu) p^k_\infty,
\end{align}
where the $p^k_\infty$ term corresponds to the $k/2$-PN order.
By expanding the 2PN result from Eq.~\eqref{eq:DeltaE_PN} the $A_{k=\{0,2,4\}}$ coefficients can be determined at every PM order.
Furthermore, in this large-$j$ limit additional PN terms have been computed.
For example, the 7PN expansion of $\mathcal{E}^{\rm 3PM}$ can be found in Eq.~(5.19) of \cite{Bini:2021gat}.
The expansion in large $j$ (at fixed $p_\infty$) introduces 1.5PN terms which were computed through 7PM in \cite{Bini:2021gat} -- see their Appendix D.
Most recently all the terms up to 3PN through 15PM have been computed Cho and collaborators \cite{Cho:2022pqy, Cho:2021onr}.

Another important limit is the parabolic limit where $e\rightarrow 1$.
For parabolic orbits the results at 1PN \cite{Blanchet:1989cu}, 2PN \cite{Bini:2020hmy}, and 3PN \cite{Cho:2022pqy} are exact.

Although the leading-order PN horizon energy flux has been calculated for circular orbits \cite{Alvi:2001mx} I am not aware of any PN calculation of the horizon flux for hyperbolic or parabolic orbits.

Above I have focused on the radiated energy, but similar results are known for the radiated angular momentum.
This was computed to 1PN order in \cite{Junker:1992kle} with a correction noted in footnote 19 of \cite{Bini:2021gat}.
Reference \cite{Bini:2021gat} also extends these results to 2PN and provides the 1.5PN terms in the large-$j$ limit.

\subsection{Post-Minkowskian results}

Since 2021 there has been rapid progress computing PM results for hyperbolic orbits.
Results for the radiated energy are summarized below.

\subsubsection{Energy radiated to infinity}

% Figures made in explore_hyperbolic_data_fixed_rmin_vary_vinf_NEW.nb
\begin{figure*}
	\includegraphics[width=0.47\textwidth]{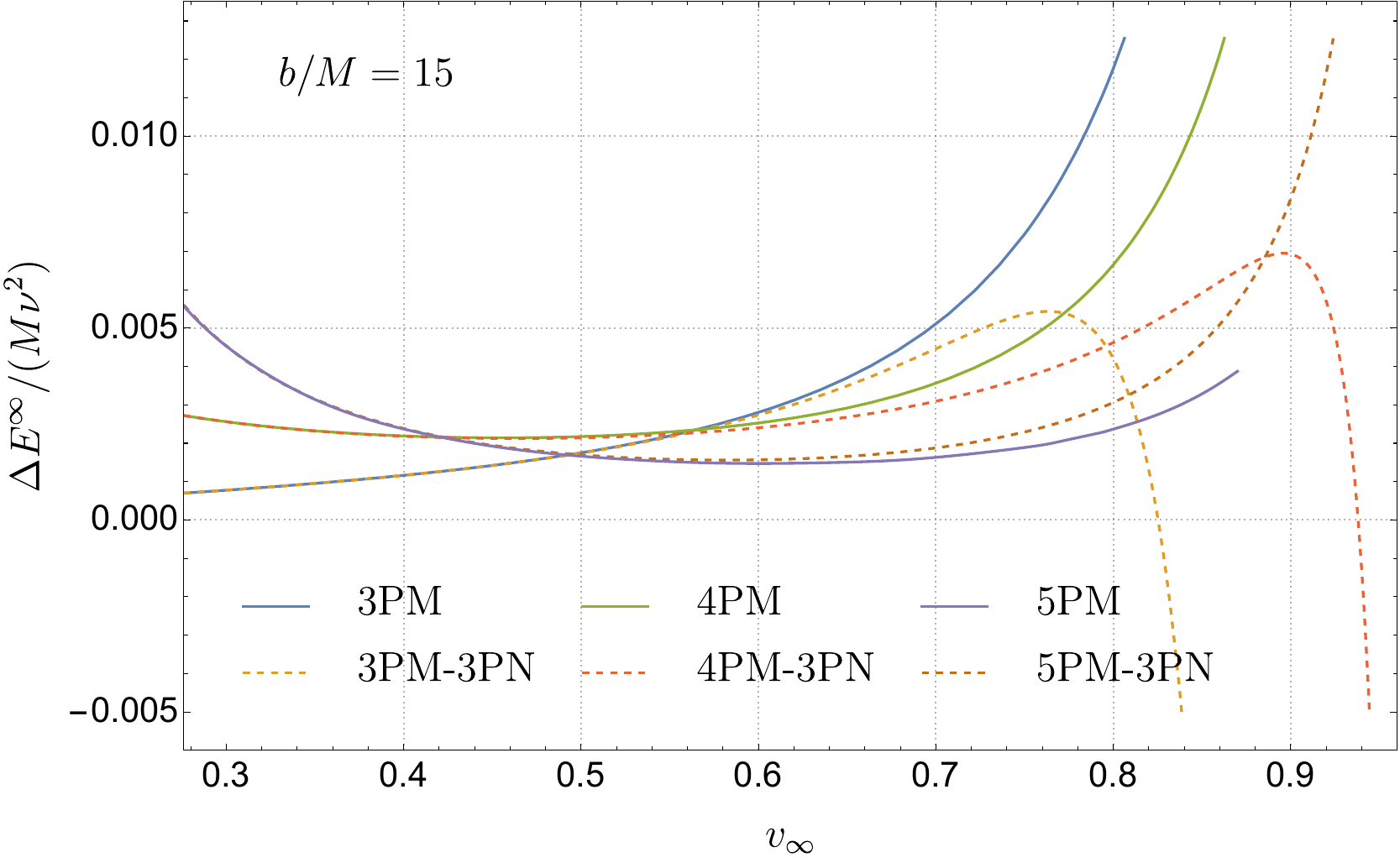}\hfill
	\includegraphics[width=0.49\textwidth]{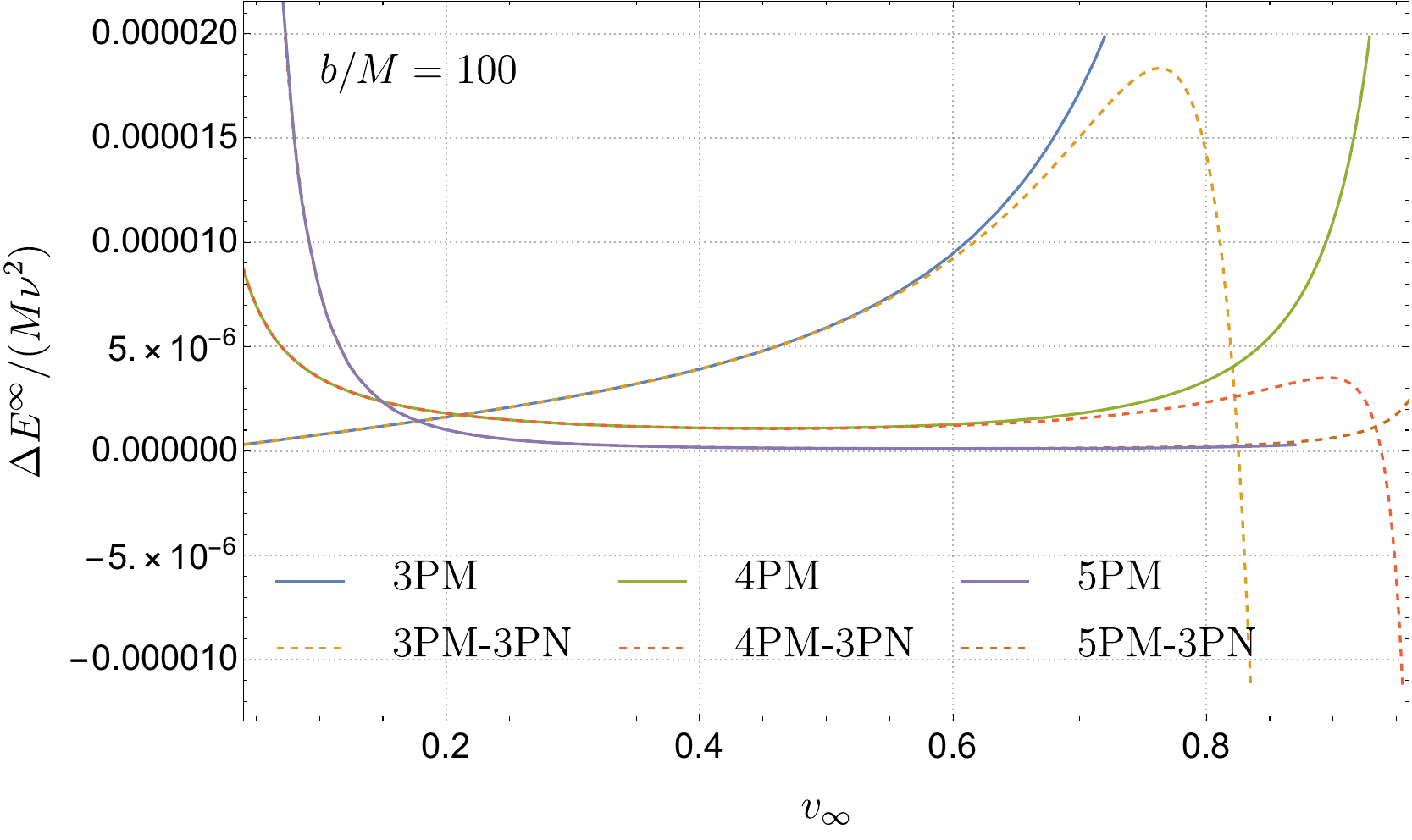}
	\caption{
		Comparison of the $\mathcal{O}(\nu^2)$ contribution to the radiated energy between PM (solid curves) and PM-3PN (dashed curves) results.
		The left plot shows the comparison in the relatively strong field with $b/M=15$.
		The right plot shows the comparison in the weak field with $b/M=100$ (this value is more typical of the values explored in this work).
		In both cases it is notable how well the PN expansion at each PM order agrees with the full PM result, even for quite large values of $\vinf$.
		In both plots the lowest value of $\vinf$ shown corresponds to the critical velocity for that value of $b$ below which the geodesic plunges into the black hole.
		Note that the weak-field criterion in Eq.~\eqref{eq:weakfield_SF} means the PM and PM-3PN results shown here will not be accurate for trajectories near the critical orbit when compared to the full numerical result. 
	}\label{fig:PMvsPM3PN}
\end{figure*}

In the PM expansion the energy radiated to infinity takes the form
% Agrees with Eq. (10) of 2101.07255 and Eq. (8) of 2210.05541
\begin{align}\label{eq:DeltaE_PM}
	\Delta E_\infty^{\rm PM} =& \frac{M^4\nu^2}{\Gamma(\nu,\gamma)}\left[\frac{\mathcal{E}^{\rm 3PM}(\gamma)}{b^3}  + \frac{\mathcal{E}^{\rm 4PM}(\gamma,\nu)}{b^4} \right. \nonumber\\
	&\left. + \frac{\mathcal{E}^{\rm 5PM}(\gamma,\nu)}{b^5} +\mathcal{O}(b^{-6})   \right].
\end{align}
%\NW{Discuss how the ``mass-polynomiality'' means that the higher-order terms do not have $\log(b)$ contributions - cite relevant Damour paper which I think is \cite{Damour:2019lcq}.}
%where $\gamma = (1-\vinf^2)^{-1/2}$ \NW{Check if this last one holds to all orders in $\nu$}.
The function $\mathcal{E}^{\rm 3PM}(\gamma)$ was first computed in Ref.~\cite{Herrmann:2021lqe}.
The 4PM term, $\mathcal{E}^{\rm 4PM}(\gamma,\nu)$ was computed in Ref.~\cite{Dlapa:2022lmu,Damgaard:2023ttc}.
The 5PM function is only known at leading order in the mass ratio; that is $\mathcal{E}^{\rm 5PM}(\gamma,\nu) = \mathcal{E}^{\rm 5PM,1SF}(\gamma) + \mathcal{O}(\nu)$ where $\mathcal{E}^{\rm 5PM,1SF}(\gamma)$ was computed in Ref.~\cite{Driesse:2024feo}.\footnote{Note that the 5PM-1SF analytic result contains special functions that can only be evaluated numerically, or can be easily expanded to extremely high PN order (the authors provided the expansion through 250PN).}
All of these results were computed within the past few years using particle physics methods \cite{Bjerrum-Bohr:2022blt, Kosower:2022yvp, Buonanno:2022pgc}.

For the leading-order, $\mathcal{O}(\nu^2)$, contribution to the radiated energy that is the focus of this work, it is interesting to note how well the PM-3PN expansion from Eq.~\eqref{eq:DeltaE_PM_PN} performs even for quite high values of $\vinf$ when compared to the full PM result of Eq.~\eqref{eq:DeltaE_PM} --- see Fig.~\ref{fig:PMvsPM3PN} for two examples.

\subsubsection{Energy radiated through the horizon}

The leading-order horizon flux was calculated in~\cite{Goldberger:2020wbx,Jones:2023ugm,Bautista:2024emt}.
It enters at 7PM order and is given by
% Agrees with Eq. (3.40) of 2310.00069 where y = u1.u2 = \gamma
\begin{equation}\label{eq:DeltaE_H_7PM}
    \Delta E_H^{\rm PM} = \frac{5\pi m_1^6 m_2^2}{16 b^7}(21 \gamma^4 - 14 \gamma^2 +1 )\sqrt{\gamma^2 -1} + \mathcal{O}(b^{-8}).
\end{equation}

\subsubsection{Radiated angular momentum and the dissipative scattering angle}

Although not the focus of the present work, for completeness I note that the 2PM and 3PM radiated angular momentum for non-spinning binaries can be found in \cite{Damour:2020tta} and \cite{Manohar:2022dea,DiVecchia:2022piu}, respectively.
Refs.~\cite{Heissenberg:2025ocy,Geralico:2025rof,Heissenberg:2025fcr} present the PN-expanded 4PM result. 
%To the best of my knowledge no one has computed the horizon absorption of the angular momentum in the PM regime.
Since this paper first appeared on the arXiv the leading order horizon absorption of angular momentum has been computed \cite{Cipriani:2026myb}.

Once both the radiated energy and angular momentum are computed it is possible to compute the dissipative correction to the scattering angle \cite{Damour:2020tta}.
This is has been computed through 4PM order \cite{Jakobsen:2023hig}.
Results for the dissipative correction to the scattering angle are known at 5PM order and leading order in the mass ratio within a PN-PM expansion \cite{Geralico:2025rof}.

\subsection{Post-Minkowskian-post-Newtonian hybrids}\label{sec:hybrid}

The effectiveness of the PN-PM expansion even at quite large values of $\vinf$ -- see Fig.~\ref{fig:PMvsPM3PN} -- suggests it could be useful to build a hybrid model that uses both PM and PN-PM results.
In this section I build such a model using all the PM results through 4PM \cite{Herrmann:2021lqe, Dlapa:2022lmu} and a subset of the known 3PN-15PM series \cite{Cho:2022pqy}.
In the model I do not include the 5PM results due to the computational cost of numerically evaluating the analytic result, and I find it sufficient to include up to the 2PN-7PM terms as beyond this there is little improvement when compared with my numerical data.
I build the model adding the 5PM to 7PM results through 2PN to the full 4PM results.
At leading order in the mass ratio the model is given explicitly by
%% Checked below against Eq. C3 of arXiv:2203.10872
\begin{align}
	\Delta &E^\infty_{\rm 4PM/2PN-7PM} = m_1^4 \epsilon^2 \left[\frac{\mathcal{E}^{\rm 3PM}(\gamma)}{b^3} + \frac{\mathcal{E}^{\rm 4PM}(\gamma,0)}{b^4}  \right. \nonumber\\
	  &+\left. \sum_{k=5}^7\frac{\mathcal{E}^{\rm kPM-1PN}(\gamma) + \mathcal{E}^{\rm kPM-2PN}(\gamma)}{b^k} \right],
\end{align}
where the $\mathcal{E}^{\rm kPM-1PN}(\gamma)$ contains the terms through 1PN at the $k$-th PM order, and $\mathcal{E}^{\rm kPM-2PN}(\gamma)$ contains the 1.5PN and 2PN terms at the $k$-th PM order.
At 5PM the terms are given explicitly by
\begin{align}
	\mathcal{E}^{\rm 5PM-1PN}(\gamma) &= p_\infty^{-3}\left(\frac{122 \pi }{5} + \frac{13831 \pi}{280} p_\infty^2\right)\\
	\mathcal{E}^{\rm 5PM-2PN}(\gamma) &= p_\infty^{-3}\left(\frac{297 \pi ^3 p_\infty^3}{20}-\frac{64579 \pi}{5040}p_\infty^4 \right)
\end{align}
At 6PM the terms are
\begin{align}
	\mathcal{E}^{\rm 6PM-1PN}(\gamma) &= p_\infty^{-5}\left(\frac{4672}{45} + \frac{142112}{315}p_\infty^2  \right)\\
	\mathcal{E}^{\rm 6PM-2PN}(\gamma) &= p_\infty^{-5}\left[ \left(\frac{9344}{45} + \frac{88576 \pi ^2}{675}\right) p_\infty^3 \right. \nonumber\\
		 							  &\quad\left.- \frac{293992}{1701}p^4_\infty \right]
\end{align}
Finally, at 7PM the terms are
\begin{align}
	\mathcal{E}^{\rm 7PM-1PN}(\gamma) &= p_\infty^{-7}\left( \frac{85 \pi }{3} + \frac{2259 \pi}{8} p_\infty^2\right)\\
	\mathcal{E}^{\rm 7PM-2PN}(\gamma) &= p_\infty^{-7}\left[\left(\frac{1579 \pi ^3}{3}-\frac{2755 \pi ^5}{64}\right) p_\infty^3  \right.  \nonumber\\
									  & \quad\left. + \frac{19319 \pi}{378}p_\infty^4\right]
\end{align}
If needed in future work it is straightforward to build hybrid models that included higher-order PN-PM terms.

\section{Hyperbolic geodesics}\label{sec:geodesics}

In the self-force approach one typically models $m_2$ as a particle orbiting the black hole $m_1$.
I will work with the usual Schwarzschild coordinates which are centered on $m_1$.
Let the coordinates of the particle be denoted by $x^\alpha_p(\tau) \equiv \{t_p,r_p,\theta_p,\varphi_p\}$ where $\tau$ is the proper time along the orbit.
I define the components of the four-velocity as $u^\alpha(\tau) = dx^\alpha_p/d\tau$.
Up to arbitrary rotations, geodesics in Schwarzschild spacetime can be uniquely specified by two constants of motion.
Once such pair follows from the time translation and rotational invariance of the spacetime: the (specific) energy and angular momentum, $\E = E_2/m_2 = \gamma$ and $\calL = L_2/m_2$, respectively.

Using these, and the normalization of the 4-velocity, $u^\alpha u_\alpha = -1$, the geodesic equations can be written in first-order form
% Agrees with Eq. 1-3 of 2209.03740
\begin{align}
	u^t 	&= \frac{\E}{f(r_p)},	\label{eq:ut}\\
	u^\phi 	&= \frac{\calL}{r_p^2},  \label{eq:uphi}\\
	(u^r)^2	&= \E^2 - \mathcal{V}_{\rm eff}(r;\calL), \label{eq:ur}
\end{align}
where $f(r_p) = 1-2m_1/r_p$ and the effective potential $\mathcal{V}_{\rm eff}(r;\calL) = f(r_p)(1+\calL^2/r_p^2)$.
Not all values of $\E$ and $\calL$ correspond to scattered orbits.
For a given $\E$, the particle will fall into the black hole if $\calL < \calL_{\rm crit}(\E)$ where the critical value is obtained by simultaneously solving $\partial_r \mathcal{V}_{\rm eff}(r;\calL) = 0$ and $\E^2 - \mathcal{V}_{\rm eff}(r;\calL) = 0$.
This gives
% Checked against Eq. 6 of 2209.03740.
\begin{align}
	\calL_{\rm crit}(\E) = m_1 \sqrt{\frac{27\E^4 + 9 \alpha \E^3 - 36 \E^2 - 8\alpha \E + 8}{2(\E^2 -1)}},
\end{align}
where $\alpha = \sqrt{9\E^2 -8}$.

The structure of the PM expansions in Eqs.~\eqref{eq:DeltaE_PM} and \eqref{eq:DeltaE_H_7PM} is clearest when the orbit is parameterized using the impact parameter and, say, $\vinf$.
The relation between $(\E,\calL)$ and $(b, \vinf)$ is given in Eq.~\eqref{eq:v_and_rel_gamma} and \eqref{eq:b_from_E_L}.
The impact parameter can also be related to the other quantities using the geodesic equations above.
Geometrically, the impact parameter is given by
\begin{align}
    b &= \lim_{\tau\rightarrow-\infty}r_p(\tau)\sin|\varphi_p(\tau) - \varphi_p(-\infty)|, \\
	  &= \lim_{\tau\rightarrow-\infty}\frac{r_p(\tau)^2u^\varphi(\tau)}{u^r(\tau)}\cos(\varphi_p(\tau) - \varphi_p(-\infty)),\\
	  &= \frac{\calL}{\sqrt{\E^2-1}} = \frac{\calL}{\vinf \E},
\end{align}
where in the second line I have used l'H\^{o}pital's rule, and the third line is arrived at by using Eqs.~\eqref{eq:uphi} and \eqref{eq:ur}.
As expected, this agrees with Eq.~\eqref{eq:b_from_E_L}.
I will denote the critical value of the impact parameter by $b_{\rm crit}(\E) = \calL_{\rm crit}(\E)/\sqrt{\E^2 - 1}$.

In order to parametrize motion along the orbit it is convenient to employ Darwin's relativistic anomaly, $\chi$.
This is defined via its relation with the orbital radius such that
\begin{align}\label{eq:rp-chi}
	r_p(\chi) = \frac{m_1 p}{1+e\cos\chi},
\end{align}
where $p$ is the semilatus rectum and $e$ is the eccentricity, and the value of $\chi$ ranges from $\chi_{\max} = \arccos(-1/e)$ to $\chi_{\min} = -\chi_{\max}$.
These constants of the motion, $(p,e)$, are defined from the roots of the cubic-in-$r_p$ equation $u^r = 0$ with
\begin{align}
	p = \frac{2r_1 r_{\min}}{m_1(r_1 + r_{\min})}, \qquad e = \frac{r_1 - r_{\min}}{r_1 + r_{\min}}
\end{align}
where $r_{\min}$ is the periastron radius which occurs at $\chi=0$.
%The three root can be computed by solving the cubic-in-$r_p$ equation $u^r = 0$.
These roots can be written explicitly as \cite{Barack:2022pde}
\begin{align}
	r_{\rm min} &= \frac{6m_1}{1-2\zeta\sin(\frac{\pi}{6}-\xi)},		\\
	r_1 		&= \frac{6m_1}{1-2\zeta\sin\left(\frac{\pi}{6}+\xi\right)},		\\
	r_2 		&= \frac{6m_1}{1+2\zeta\cos\xi},
\end{align}
where $\zeta = \sqrt{1-12m_1^2/\calL^2}$ and
\begin{align}
	\xi = \frac{1}{3}\arccos\left(\frac{1+(36-54\E^2)m_1^2/\calL^2}{\zeta^3}\right).
\end{align}

By noting that the field will be strongest when the particle is at $r_{\rm min}$ the weak-field condition in \eqref{eq:weakfield} becomes $m_1/r_{\rm min} \ll 1$.
Reference \cite{Barack:2022pde} showed that $r_{\rm min} = b(1 - m_1/(\vinf^2 b)) + \mathcal{O}(b^{-1})$ and hence the weak-field condition will hold when both $b \gg 1$ and 
\begin{align}\label{eq:weakfield_SF}
    \frac{m_1}{\vinf^2 b} \ll 1.
\end{align}

The relation between $(p,e)$ and $(\E,\calL)$ is given by \cite{Cutler:1994pb}
\begin{align}
	\E^2 &= \frac{(p-2)^2 - 4e^2}{p(p-3-e^2)}\\
	\calL^2 &= \frac{p^2 m_1^2}{p-3-e^2}  
\end{align}

The evolution of $t_p$ and $\varphi_p$ with respect to $\chi$ can be derived using Eqs.~\eqref{eq:ut}, \eqref{eq:uphi} and \eqref{eq:rp-chi} to give
\begin{align}
	\frac{dt_p}{d\chi} &= \frac{m_1p^2}{(p-2-2e\cos\chi)(1+e\cos\chi)^2}\sqrt{\frac{(p-2)^2-4e^2}{p-6-2e\cos\chi}} \label{eq:dtdchi}\\
	\frac{d\varphi_p}{d\chi} &= \sqrt{\frac{p}{p-6-2e\cos\chi}}	\label{eq:dphidchi}
\end{align}
Both can be integrated with the result given in terms of elliptic integrals.
The function $t_p(\chi)$ is given in Eq.~(B7) of \cite{VanDeMeent:2018cgn}. 
%\NW{Should I use this in the code?}
For $\varphi_p$ we have
\begin{align}
	\varphi_p(\chi) = k\sqrt{\frac{p}{e}} F\left(\frac{\chi}{2},-k^2\right),
\end{align}
where $k^2 = 4e/(p-6-2e)$ and $F(\phi,z) = \int_0^\phi (1-z\sin\theta)^{-1/2}\,d\theta$.

\section{Self-force calculation}\label{sec:SF_calc}

The goal of this work is to compute the total energy radiated during a scattering event.
In order to compute the radiation I make use of the RWZ formalism \cite{Regge:1957td,Zerilli:1970se}.
Assuming that $\epsilon \equiv m_2/m_1 \ll 1$ then the metric of the two-body system, $g_{\alpha\beta}$ is dominated by the metric of the primary mass, $\bar{g}_{\alpha\beta}$, so that we have $g_{\alpha\beta} = \bar{g}_{\alpha\beta} + \epsilon h_{\alpha\beta} + \mathcal{O}(\epsilon^2)$.
Making use of the spherical symmetry of the background Schwarzschild metric the perturbation can be decomposed onto a basis of spin-weighted spherical harmonics. 
An analysis of the gauge freedom of the perturbation reveals that one can choose to set certain components of the metric to zero which puts the metric perturbation into the Regge-Wheeler gauge \cite{Martel:2005ir}.
Within this gauge master functions, $\Psi_{\lm}$, can be derived from which the radiated energy and angular momentum can be computed.
Rather than use the original RWZ variables it is now common to use the Zerilli-Moncrief (ZM) \cite{Moncrief:1974am} and Cunningham-Price-Moncrief (CPM) \cite{Cunningham:1978zfa} formulations which allow for reconstruction of the metric perturbation from the master function in the time domain without carrying out any integration \cite{Martel:2005ir, Hopper:2010uv}.

Asymptotically, in the wave zone, the two gravitational-wave polarizations are related to the ZM and CPM master functions via \cite{Pound:2021qin}
\begin{align}
	h_+ - i h_\times = \frac{1}{2r}\sum_{\lm}\sqrt{\frac{(l+2)!}{(l-2)!}}\left(\Psi^{ZM}_{\lm} - i \Psi^{CPM}_{\lm} \right)  {}_{-2}Y_{\lm}.
\end{align}
Substituting this into the Isaacson stress-energy tensor \cite{Isaacson:1968zza} one can derive the flux of energy and angular momentum.
Upon integrating along the scattered orbit the total radiated energy and angular momentum are given by~\cite{Hopper:2010uv}
% Agrees with Eq. (11) of 1706.02791
\begin{align}
	\Delta E^\infty &= \sum_{\lm} \int_{-\infty}^\infty \frac{1}{64\pi}\frac{(l+2)!}{(l-2)!}|\dot{\Psi}_\lm|^2\, du,		\\
	\Delta L^\infty &= \sum_{\lm} \int_{-\infty}^\infty \frac{im}{64\pi}\frac{(l+2)!}{(l-2)!} \dot{\Psi}_\lm \bar{\Psi}_\lm \, du,
\end{align}
where $u = t - r_*$ with $r_* = r+ 2M\log(r/2M-1)$ as the tortoise coordinate, an overdot denotes differentiation with respect to $t$, an overbar denotes complex conjugation, $\Psi_\lm = \Psi_\lm^{CPM}(u,r_*\rightarrow \infty)$ for $\ell+m=\rm{odd}$, and $\Psi_\lm = \Psi_\lm^{ZM}(u,r_* \rightarrow \infty)$ for $\ell+m=\rm{even}$.
Analogous results exist for the flux through the horizon \cite{Hopper:2010uv}.

The ZM and CPM master functions are governed by the following ordinary differential equations
% Agrees with Eq. (3) of 1706.02791
\begin{align}\label{eq:RWZ_TD}
    \mathcal{L}_\ell \Psi^{\rm ZM/CPM}_\lm(t,r) = S^{\rm ZM/CPM}_{\lm}(t,r),
\end{align}
where the operator $\mathcal{L}_\ell$ is given by
% Agrees with Eq. (3) of 1706.02791
\begin{align}\label{eq:TD_operator}
	\mathcal{L}_\ell =  -\frac{\partial^2}{\partial t^2} + \frac{\partial^2}{\partial r_*^2} - V_\ell(r).
\end{align}
The potential $V_\ell(r)$ is $(l+m)$-parity dependent.
In the odd sector one uses the Regge-Wheeler potential
% Agrees with Eq. (C17) of 1006.4907
\begin{align}
	V^{\rm odd}_\ell(r) = \frac{f(r)}{r^2}\left[\ell(\ell +1) - \frac{6m_1}{r}\right],
\end{align}
where $f(r) = 1-2m_1/r$ and in the even sector one uses the Zerilli potential
% Agrees with Eq. (C9) of 1006.4907
\begin{align}
V^{\rm even}_\ell(r) = \frac{f(r)}{r^2\Lambda(r)^2}\left[ 2\lambda^2\left(\lambda + 1 +\frac{3m_1}{r}\right) \right. \nonumber \\
 \left. + \frac{18m_1^2}{r^2}\left(\lambda + \frac{m_1}{r}\right) \right],
\end{align}
where $\Lambda(r) \equiv \lambda + 3m_1/r$ and $\lambda \equiv (\ell+2)(\ell -1)/2$.
 
In general the source in Eq.~\eqref{eq:RWZ_TD} has the form \cite{Martel:2003jj}
\begin{align}\label{eq:source_TD}
	S_\lm(t,r) = G_\lm(t,r) \delta(r - r_p(t)) + F_\lm(t,r) \delta'(r-r_p(t))
\end{align}
where hereafter I have dropped the ZM/CPM labels for brevity.
Due to the distributions in the source it only has support at $r=r_p(t)$.
It will be useful later to write the source in the `fully evaluated form' as
\begin{align}\label{eq:source_TD_fully_evaluated}
	S_\lm(t,r) = \tilde{G}_\lm(t) \delta(r-r_p(t)) + \tilde{F}_\lm(t)(r-r_p(t))
\end{align}
where
\begin{align}
	\tilde{G}_\lm(t) &= \left[G_\lm(t,r) - \partial_r F_\lm(t,r)\right]_{r=r_p(t)}, \\
	\tilde{F}_\lm(t) &= \left[F_\lm\right]_{r=r_p(t)};
\end{align}
see Appendix A of Ref.~\cite{Hopper:2010uv} for details.
The explicit form of the $\tilde{G}_\lm(t)$ and $\tilde{F}_\lm(t)$ functions for the ZM and CPM formulation is given in Appendix B of \cite{Hopper:2010uv}.

A variety of works have solved the 1+1 RWZ equations \eqref{eq:RWZ_TD} for bound orbits \cite{Martel:2003jj,Field:2009kk,DaSilva:2023xif} but there are additional challenges when considering hyperbolic orbits.
These stem from the (typically) finite computational domain and interaction with so-called junk radiation which occurs when the initial conditions do not satisfy the field equations \cite{Long:2021ufh}.
%\NW{I think there is a paper that looks a the ultra relativistic limit and finds it numerically challenging due to interaction with the junk radiation.}

An alternative approach is to further decompose the master function into the frequency domain.
At first glance this approach might not seem particularly attractive as it is well known that frequency-domain calculations become more computationally demanding as $e\rightarrow1$.
Fortunately, once $e\ge1$ the approach becomes practical again as was shown by Hopper \cite{Hopper:2017iyq} and further illuminated by Whittall and Barack \cite{Whittall:2023xjp}.

I proceed by taking the Fourier transform of the master functions and the source as
%Note this is different from Seth's definition in Eq. (10) of 1706.02791 as I don't include the 1/(2\pi). I follow Eq. (47) of 2305.09724
\begin{align}
    \Psi_\lm(t,r) &= \int^\infty_{-\infty} X_\lmo(r) e^{-i\omega t}\, d\omega, \\
    S_\lm(t,r) 	  &= \int^\infty_{-\infty} Z_\lmo(r)e^{-i\omega t}\,d\omega.
\end{align}
Using these we get an ordinary differential equation for each $\lmo$-mode of the perturbation
\begin{align}\label{eq:RWZ}
    \mathcal{L}_{\ell\omega}X_\lmo(r) = Z^{\rm RW/Z}_{\ell m \omega}(r),
\end{align}
where the frequency-domain operator $\mathcal{L}_{\ell\omega}$ is given by
\begin{align}
	\mathcal{L}_{\ell\omega} = \frac{d^2}{dr_*^2} + \omega^2 - V^{\rm{RW/Z}}_\ell(r).
\end{align}
%The explicit form of the frequency domain sources $Z^{\rm RW/Z}_{\ell m \omega}$ can be found in Appendix \ref{apdx:source_terms}.

The general solution to this inhomogeneous equation \eqref{eq:RWZ} can be written using the method of variation of parameters
\begin{align}\label{eq:X}
    X_\lmo(r) = C^+_\lmo(r)\hat{X}^+_\lmo(r) + C^-_\lmo(r)\hat{X}^-_\lmo(r),
\end{align}
where $\hat{X}^\pm_\lmo$ are two linearly independent homogeneous solutions to Eq.~\eqref{eq:RWZ}.
In order to select the retarded solution the boundary conditions for the homogeneous solution are given by
\begin{align}\label{eq:X_BCs}
	\hat{X}_\lmo^\pm(r_* \rightarrow  \pm \infty) = e^{\pm i \omega r_*}
\end{align}
where, without loss of generality, we have unit normalized the asymptotic homogeneous solutions, i.e., $\lim_{r_* \rightarrow \pm \infty}|\hat{X}^{\pm}_\lmo(r_*)| = 1$.
The weighting coefficients, $C^\pm_\lmo(r)$ are given by
\begin{align}
	C^\pm_\lmo(r) = \pm\frac{1}{W_\lmo}\int_{r_{\min}}^r \frac{X^\mp_\lmo(r') Z_\lmo(r')}{f(r')}\,dr',
\end{align}
where $W_\lmo$ is the (constant) Wronskian formed from the homogeneous solutions.

For the calculation of the asymptotic fluxes we only need the asymptotic values of the $C^\pm_\lmo$ which, after substituting the explicit form of the source, is given by
% Agrees with Eq. (3.15) of 1006.4907
\begin{align}\label{eq:weighting_coeffs}
    &C^\pm_\lmo = \frac{1}{W_\lmo}\int^\infty_{-\infty}\left[\frac{1}{f_p}\hat{X}^\mp_\lmo(r_p) \tilde{G}_\lm(t)\right. \\
    &+\left.\left(\frac{2M}{r_p^2 f_p^2}\hat{X}^\mp_\lmo(r_p) - \frac{1}{f_p}\frac{d\hat{X}^\mp_\lmo(r_p)}{dr}\right)\tilde{F}_\lm(t)\right]e^{i\omega t}\,dt. \nonumber
\end{align}
where $f_p = f(r_p)$. 
Once the weighting coefficients are known the total radiated energy and angular momentum are given by
\begin{align}
	\Delta E^\pm = \sum_{l=2}^{\infty}\sum_{m=-l}^l \frac{1}{64\pi}\frac{(l+2)!}{(l-2)!}\int_{-\infty}^\infty \omega^2 |C_\lmo^\pm|^2\,d\omega \label{eq:DeltaE} \\
	\Delta L^\pm = \sum_{l=2}^{\infty}\sum_{m=-l}^l \frac{m}{64\pi}\frac{(l+2)!}{(l-2)!}\int_{-\infty}^\infty \omega |C_\lmo^\pm|^2\,d\omega,	  \label{eq:DeltaL} 
\end{align}
where $\Delta E^+ \equiv \Delta E^\infty$ is the energy radiated to infinity, $\Delta E^-  \equiv \Delta E^H$ is the energy absorbed by the black hole, and similarly for the angular momentum.
%\NW{Is there a factor of $2\pi$ missing here? See Eq.~(12) of \cite{Hopper:2017qus}. Ans: depend on the definition of the inverse Fourier transform.}

As was pointed out by Hopper \cite{Hopper:2017iyq}, the integral in Eq.~\eqref{eq:weighting_coeffs} converges very slowly (if it converges at all) if you use the standard CPM and ZM variables.
Fortunately, Hopper also showed how to derive new master functions which obey the same homogeneous ordinary differential equation (ODE) but for which the sources fall off faster at large radius and thus the integrand converges more rapidly.\footnote{Whittall and Barack \cite{Whittall:2023xjp} developed an alternative approach to increase the rate of convergence of the integrand using integration by parts which I have not investigated in the RWZ.
%\NW{could probably also use a puncture field, like we do at second-order.}
}
Hopper showed that each ``higher-order master function'' is related to the previous one via
\begin{align}\label{eq:Psi_n} 
	\Psi^{(n+1)}_\lm(t,r) = \dot{\Psi}^{(n)}_\lm + \frac{\dot{r}_p \E^2}{f_p^2 V_p^2}\tilde{F}_\lm^{(n)}\delta(r-r_p(t)),
\end{align}
where the $n$ index denotes the order of the source.
By construction, when one acts on $\Psi_\lm^{(n+1)}$ with $\mathcal{L}_{\ell}$ the coefficient of the $\delta''(r-r_p(t))$ cancels and the resulting source takes the same form as Eq.~\eqref{eq:source_TD_fully_evaluated}, i.e.,
\begin{align}\label{eq:RWZ_TD_higher_order}
	\mathcal{L}_{\ell} \Psi^{(n+1)}_\lm = \tilde{G}_\lm^{(n+1)}(t)\delta(r-r_p(t)) + \tilde{F}_\lm^{(n+1)}(t)\delta'(r-r_p(t)).
\end{align}
Defining $\tilde{G}_{\lm}^{(0)}(t) \equiv \tilde{G}_{\lm}(t)$ and $\tilde{F}_{\lm}^{(0)}(t) \equiv \tilde{F}_{\lm}(t)$ and using Eqs.~\eqref{eq:Psi_n} and \eqref{eq:RWZ_TD_higher_order} one can derive explicit forms for  $\tilde{G}_\lm^{(n+1)}(t)$ and $\tilde{F}_\lm^{(n+1)}(t)$.
Hopper provides an explicit recurrence relation between the $n$-th and $(n+1)$-th sources -- see his Eqs.~(25) and (26) in Ref.~\cite{Hopper:2017iyq}.

As Eq.~\eqref{eq:RWZ_TD_higher_order} has the same form as Eq.~\eqref{eq:RWZ_TD} when one transforms to the frequency domain the resulting equations have the same form as those described above.
In particular, we have
\begin{align}\label{eq:RWZ_higher_order}
	\mathcal{L}_{\ell\omega} X_\lmo^{(n)}(r) &= Z^{(n)}_{\ell m \omega}(r),
\end{align}
and
\begin{align}\label{eq:X_higher_order}
	X_\lmo^{(n)}(r) &= C^{(n)+}_\lmo(r)\hat{X}^+_\lmo(r) + C^{(n)-}_\lmo(r)\hat{X}^-_\lmo(r),
\end{align}
where the weighting coefficients $C^{(n)\pm}_\lmo(r)$ are computed via Eq.~\eqref{eq:weighting_coeffs} with the replacements $C^{\pm}_\lmo(r)\rightarrow C^{(n)\pm}_\lmo(r)$, $\tilde{G}_{\lm}(t) \rightarrow \tilde{G}^{(n)}_{\lm}(t)$, and $\tilde{F}_{\lm}(t) \rightarrow \tilde{F}^{(n)}_{\lm}(t)$.
The form of Eqs.~\eqref{eq:RWZ_higher_order} and \eqref{eq:X_higher_order} is particularly useful as it allows one to recycle computational infrastructure set up to solve for the ZM or CPM variables.
For example, as the operator is the same in Eqs.~\eqref{eq:RWZ_higher_order} and \eqref{eq:RWZ}, the homogeneous solutions in Eq.~\eqref{eq:X_higher_order} are the same as those in Eq.~\eqref{eq:X}.

Although higher-order sources fall off more rapidly, and thus the integral in Eq.~\eqref{eq:weighting_coeffs} converges faster, the complexity of the source grows exponentially and so for practical calculations there is a trade-off between more a rapidly convergent integrand and the computational cost to evaluate the source; see Fig.~2 of Ref.~\cite{Hopper:2017iyq} for a quantitative analysis.
For the present calculation, I find it practical to work with $\Psi^{(2)}_{ZM}$ and $\Psi^{(1)}_{CPM}$ which ensures that the sources fall off as $r_p^{-3}$.
The explicit form of the sources for these fields can be found in Appendix B of Ref.~\cite{Hopper:2017iyq}.

From Eq.~\eqref{eq:Psi_n} we see that, away from the worldline,  each $\Psi_\lmo^{(n+1)}$ is just the time derivative of the $\Psi^{(n)}_\lmo$.
It is thus straightforward to relate the asymptotic weighting coefficients of the higher-order master functions to the originals via
\begin{align}
	C^{(0)\pm}_{lm\omega} = \frac{C^{(n)\pm}_{lm\omega}}{(-i\omega)^n}.
\end{align}
The radiated energy and angular momentum can be computed using the standard formula in Eqs.~\eqref{eq:DeltaE} and \eqref{eq:DeltaL}.

\section{Numerical implementation, error analysis, and validation}\label{sec:numerical_method}

%Plot made in RWZ_hyperbolic.nb
\begin{figure}
    \centering
    \includegraphics[width=0.95\linewidth]{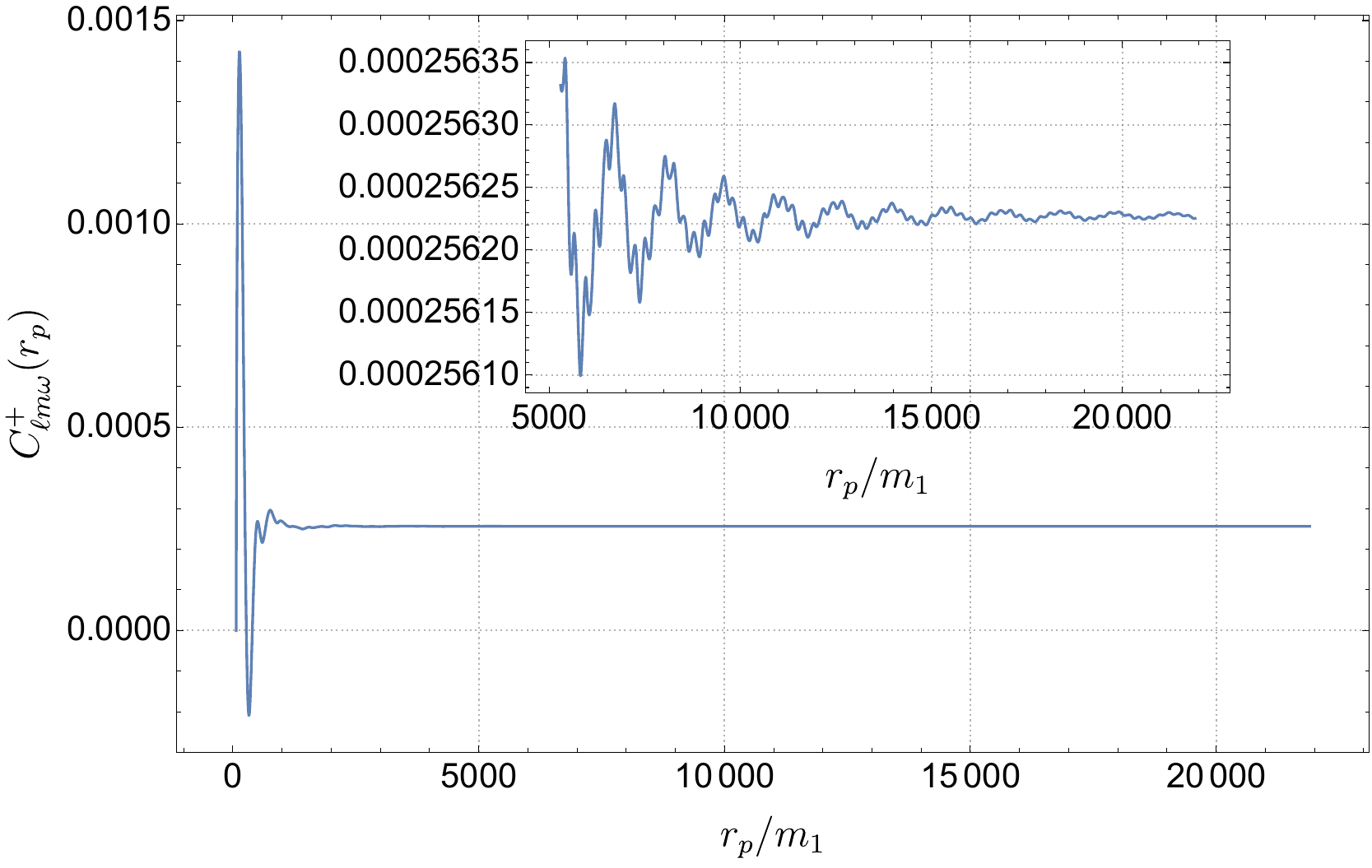}
    \caption{Convergence of the weighting coefficient for the up solution for the $l=2, m=2$ mode with $\omega=0.01$ and $b=70$, $\vinf=0.7$.
	Although the integral appears to converge quite rapidly, the inset shows that the integrand is oscillating rapidly.
	Resolving all these oscillations is a significant computational bottleneck.
	I use the magnitude of the oscillations near the largest value of $r_p/m_1$ as an estimate on the error in $C^\pm_\lmo$.
	}
    \label{fig:cplus_convergence}
\end{figure}

In this section I detail the numerical approach I use to compute the weighting coefficients by evaluating the integral in Eq.~\eqref{eq:weighting_coeffs} (using the higher-order sources).
All the calculations are made using the \texttt{Mathematica} software package.

First I require the homogeneous solutions, $\hat{X}^\pm_\lmo$, to the RWZ equations.
The asymptotic boundary conditions at future null infinity and the event horizon are given by Eq.~\eqref{eq:X_BCs}.
One way to enforce these boundary conditions in a numerical scheme is to switch to hyperboloidal, compactified coordinates where using spectral methods the boundary conditions can be placed exactly on future null infinity and the event horizon \cite{PanossoMacedo:2022fdi,Leather:2024mls}.
In this work I opt to use the more traditional approach and solve the equations on $t$-slicing, as they are written in Sec.~\ref{sec:SF_calc}.
For this I have to place the numerical boundaries at a finite radius, and the boundary conditions are supplied by performing series expansions of the asymptotic field.
For example, a suitable expansion and resulting recurrence relation for $X^+_\lmo(r)$ at large $r$ can be found in Appendix D of Ref.~\cite{Hopper:2010uv}.
In practice, for my implementation I extracted the boundary condition code for both $X^+_\lmo(r)$ and $X^-_\lmo(r)$ from the \texttt{ReggeWheeler} package of the Black Hole Perturbation Toolkit \cite{BHPToolkit}.

%Left panel from explore_hyperbolic_data_fixed_vinf_vary_b_high_v.nb
%Rgith panel from explore_hyperbolic_data_fixed_vinf_vary_b.nb
\begin{figure*}
    \centering
    \includegraphics[width=0.48\linewidth]{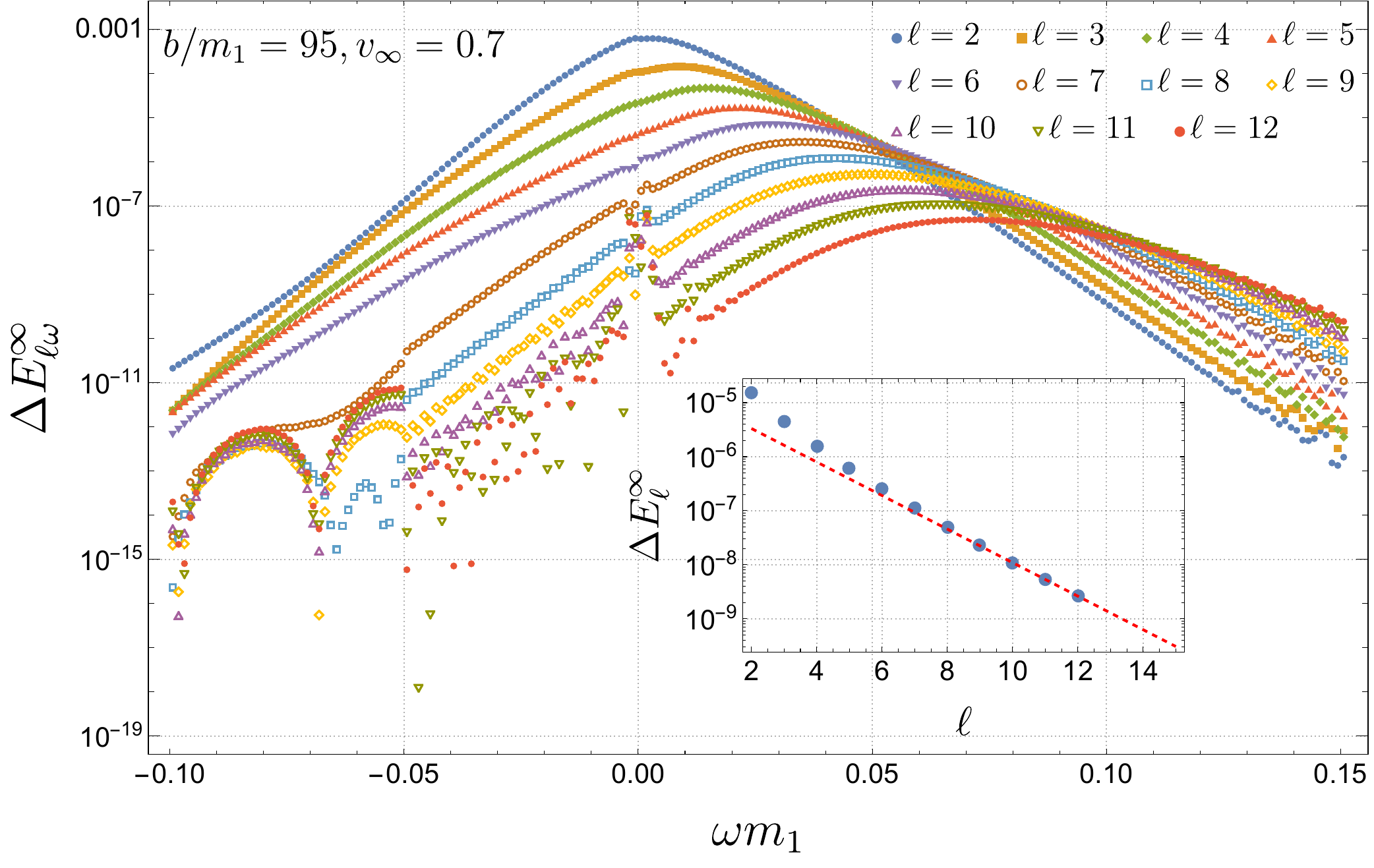}\hfill
	 \includegraphics[width=0.49\linewidth]{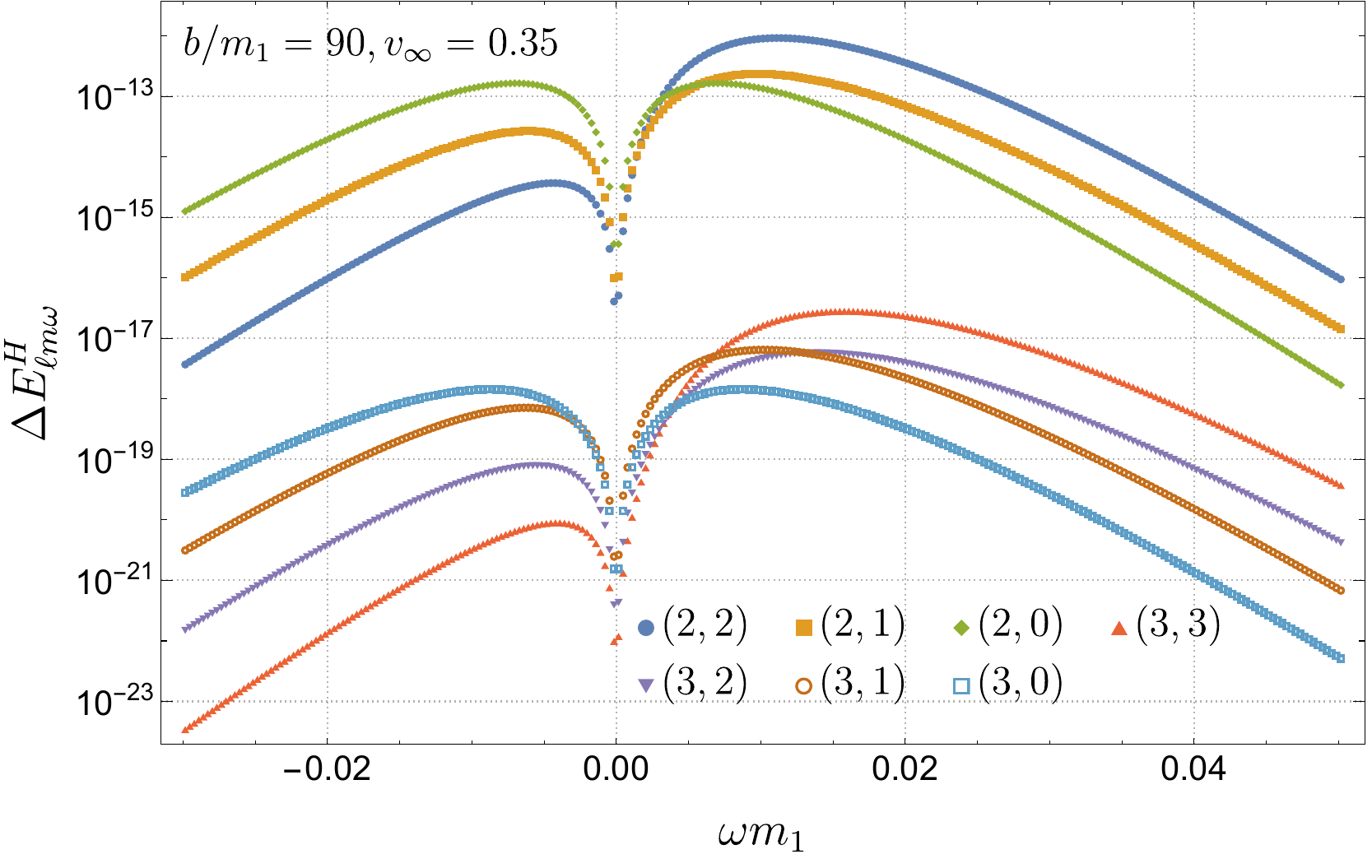}
    \caption{
		(Left panel) The spectrum of the total radiated energy for each value of $\omega m_1$ for a variety of $\l$-modes (summed over $m$) for a hyperbolic orbit with $b/m_1=95$ and $\vinf=0.7$.
		(Inset) The contribution to the $\ell$-modes of $\Delta E^\infty$ (after integrating over $\omega$).
		For large $\ell$ the contribution drops off exponentially.
		The red, dashed line shows an exponential fit to the modes between $\ell=8$ and $\ell=12$.
		(Right panel) Contribution to the total energy absorbed by the horizon for each value of $\omega m_1$ for a variety of $\lm$-modes for a hyperbolic orbit with $b/m_1=90$ and $\vinf=0.35$.
		For the weak-field orbits considered in this work the $\ell=2$ modes dominate so much that all the other modes can be ignored.
		}
    \label{fig:spectrum}
\end{figure*}
 
The calculation of the weighting coefficients for given values of $(\ell, m, \omega)$ proceeds with the following steps:
\begin{enumerate}
	\item 	I compute the numerical boundary conditions for $X^+_\lmo(r)$ at $r = r_\text{out}$, and similarly for $X^-_\lmo(r)$ at $r=r_\text{in}$.
			The outer boundary location, $r_\text{out}$, is adjusted to ensure that the boundary is in the wave zone where the boundary condition expansion will converge.
			For the inner boundary location I find it is always sufficient to set $r_\text{in}/m_1 = 2 + 10^{-5}$.
	\item 	I numerically integrate the homogeneous version of Eq.~\eqref{eq:RWZ_higher_order} from the numerical boundary conditions to $r=r_\text{min}$.
			The integration is carried out using \texttt{Mathematica}'s \texttt{NDSolve} function and I store the values of $\hat{X}^\pm_{\lmo}$ and their radial derivatives at $r=r_\text{min}$.
	\item 	In Sec.~\ref{sec:geodesics} I have parameterized the orbital motion by the Darwin parameter, $\chi$, and thus it is useful to change the integration variable in the source region from $r$ to $\chi$ by using Eq.~\eqref{eq:rp-chi} to compute $dr_p/d\chi$.
			Similarly, I change the integration variable from $t$ to $\chi$ in Eq.~\eqref{eq:weighting_coeffs} using $dt/d\chi$ from Eq.~\eqref{eq:dtdchi}.
			By defining $\hat{Y}^\pm_\lmo \equiv d\hat{X}^\pm/dr$ I write the field equation in first-order form.
			I then form a vector of equations as $[\hat{X}^{+'}_{\lmo}\,\, \hat{Y}^{+'}_{\lmo}\,\, \hat{X}^{-'}_{\lmo}\,\, \hat{Y}^{-'}_{\lmo}\,\, C^{+'}_{\lmo}\,\, C^{-'}_{\lmo}]$ where a prime denotes differentiation with respect to $\chi$.
			I take the initial conditions for the homogeneous fields to be the values I stored at $r=r_{\text{min}}$ (equivalently $\chi=0$) and I set $C^{\pm'}_{\lmo}(\chi=0) = 0$.
			I then numerically integrate this set of coupled equations from $\chi = \chi_\text{min} + \Delta\chi$ to $\chi = \chi_\text{max} - \Delta\chi$ where $\Delta\chi$ is introduced to make the limits of the integral finite (with respect to $r$).
			The precise value of $\Delta\chi$ depends on the parameters of the scattering orbit, as well as the mode indexes $(\ell,m,\omega)$.
			In practice I manually select the value of $\Delta\chi$ after experimentation with a few modes.
\end{enumerate}
The evaluation of the weighting coefficients using the prescription above is numerically demanding as the integrand of \eqref{eq:weighting_coeffs} is highly oscillatory.
Figure \ref{fig:cplus_convergence} gives an example of the rate of convergence of $C^+_\lmo$.
I find that for each value of $(\ell,m,\omega)$ it typically takes a few 10s of seconds to compute the $C^\pm_\lmo$.

In order to compute the radiated energy and angular momentum the next step is to evaluate Eqs.~\eqref{eq:DeltaE} and \eqref{eq:DeltaL}.
In practice, the integrands in these equations can only be sampled at a discrete set of frequency values over a finite range of frequencies, and the mode sum can only be carried out of a finite number of modes. 
In order to discuss the four main sources of error which effect the final results it is useful to refer to Fig.~\ref{fig:spectrum}.

\textit{Error in the computation of the $C^\pm_\lmo$.} 
This error stems from two sources.
One follows from the finite step size of the ODE solver, though with adaptive stepping this is easily controlled to be below other sources of error.
The dominant error comes from the truncation of the integral in Eq.~\eqref{eq:weighting_coeffs}.
In general I find that I need to extend the integration domain further for large values of $|\omega m_1|$ and also for $m_1 \omega \sim 0$.
The effect of not integrating sufficiently far is clearly visible in the left panel of Fig.~\ref{fig:spectrum} where noise appears for high-$\ell$ modes in these two regions.
This noise can be further reduced by extending the domain of integration, though this comes at significant computational cost and thus I avoid where possible.
For each mode I can use the magnitude of the oscillations present in the final portion of the integration, e.g., the inset of Fig.~\ref{fig:cplus_convergence}, to estimate the error in the final result.
I find this estimate gives results that are broadly consistent with the level of noise observed in the mode data.
In future iterations of the code it may be useful to estimate this error on the fly and extend the integration domain out as far as is necessary to reach some prescribed error threshold.
For now, if I find the error is becoming too large I manually decrease the value of $\Delta\chi$ for the effected $\lm$-modes and range of $\omega$ values.
I note that although noise is still clearly present in the left panel of Fig.~\ref{fig:spectrum} I have checked that the contribution from the regions and modes where the noise is present does not effect the comparisons I present in the next section.
It is also worth noting in the left panel of Fig.~\ref{fig:spectrum} the location where the value of $\Delta\chi$ changes is clearly visible, for example around $\omega m_1 = -0.05$.

\textit{Error from the discretization of the frequency spectrum.} 
Practically I can only sample the integrand in Eqs.~\eqref{eq:DeltaE} and \eqref{eq:DeltaL} on a finite grid of $\omega$ values.
For the weak-field orbits I consider in this work, I find it is sufficient to sample the $C^\pm_\lmo$ with spacing $m_1\Delta\omega \in (0.003125,0.00125)$.
I then carry out the integration using Simpson's method.
As a conservative estimate of the error I can half the number of sample points and carry out the integration again.
By comparing the results with different sample spacing I can ensure the highest sampling rate is sufficiently dense that the discretization of the spectrum is a subdominant source of error.
In future iterations of the code it might be possible to use the results from multiple sampling rates to extrapolate to an effective infinitely dense sampling via, e.g., Richardson extrapolation.

\textit{Error from the truncation of the frequency spectrum.} 
Formally the integrals in Eqs.~\eqref{eq:DeltaE} and \eqref{eq:DeltaL} must be carried out over all frequency values.
Fortunately, for large values of $|m_1\omega|$ the contributions to the radiation drop off exponentially \cite{Kovacs:1978eu, Hopper:2017qus}.
Thus, I just need to sample the frequency spectrum in a sufficiently large region around the peak.
To find the size of the region I carry out a very coarse sampling of a wide region to get the general shape of the spectrum.
From this I can then decide on the region to densely sample.
The left panel of Fig.~\ref{fig:spectrum} shows that the peak of each $\ell$-mode moves to the higher frequencies as $\ell$ increases.
For each $\ell$-mode the peak roughly occurs at $\omega = \ell \Omega_\varphi^\text{circ}$ where $\Omega_\varphi^\text{circ} = \sqrt{m_1/r_\text{min}^3}$ is the orbital frequency of a circular orbit with a radius equal to the periastron radius of the hyperbolic orbit.
This approximation becomes more accurate as the orbit approaches the critical orbit where ``zoom-whirl'' behaviour \cite{Glampedakis:2002ya} sets in and the radiation becomes dominated by the nearly circular, near-periastron whirl portion of the orbit.
It is interesting to note that secondary peaks in the spectrum can occur due to quasinormal mode excitation of the larger black hole -- see Fig.~5 in Ref.~\cite{Whittall:2023xjp} for an example in the scalar-field case.
There they observe that the secondary peaks are usually quite suppressed even for strong-field orbits and I do not find any evidence for them in the weak-field orbits considered in this work.

\textit{Error from the truncation of the sum over the spherical harmonic modes.} 
Formally, when evaluating Eqs.~\eqref{eq:DeltaE} and \eqref{eq:DeltaL} one must sum over all $(\ell,m)$-modes.
Fortunately, the contributions from the high-$\ell$-modes drop off exponentially and so I can truncate at some suitable $\ell_{\max}$.
To ensure that I am not missing a significant contribution from the uncomputed modes, I fit the last few $\ell$-modes I have available to a function of the form $a e^{-b\ell}$, where $a$ and $b$ are constants to be determined.
I then integrate this fit from $\ell=\ell_{\max}+1$ to $\infty$ and add more $\ell$-modes if this estimate is larger than a threshold I set for the calculation.
An example of the $\ell$-mode contribution to $\Delta E^\infty$ and the associated fit is shown in the inset of the left panel of Fig.~\ref{fig:spectrum}.
In general more $\ell$-modes are needed for strong-field orbits, and for orbits with larger $\vinf$.
In practice, for each $\ell$ mode it is not always necessary to compute all the $m$-modes.
For example, in the left panel of Fig.~\ref{fig:spectrum} all the $m$-modes are computed up to $\ell=6$ and beyond this I just compute a subset of $m$-modes (e.g., for $\ell=7$ I included $m=\{7,6,5\}$ and for $\ell=12$ I only include the dominate $m=12$ mode).
At the horizon the picture is very different and the $\ell=2$ modes dominate -- see the right panel of Fig.~\ref{fig:spectrum}.
As such I only include the $\ell=2$ modes in the data used for later comparisons.

% \begin{figure}
%     \centering
%     \includegraphics[width=0.95\linewidth]{figures/DeltaE_H_lmo_b90_vinf_0.35.pdf}
%     \caption{
% 		Contribution to the total energy absorbed by the horizon for each value of $\omega m_1$ for a variety of $\lm$-modes for a hyperbolic orbit with $b=90$ and $\vinf/c=0.35$.
% 		For the weak-field orbits considered in this work the $\ell=2$ modes dominate so much that all the other modes can be ignored.
% 		}
%     \label{fig:modesH}
% \end{figure}

One way to validate the numerical results is to compare them with previously published data.
The details of these comparisons can be found in Appendix \ref{apdx:parabolic} where I also present a comparison with PN results for parabolic orbits.

Another way to validate the results of this RWZ calculation would be to carry out the calculation in another gauge.
Hopper discussed how his approach could also be used to make the calculation in the Lorenz gauge tractable \cite{Hopper:2017iyq}.
As a check I implemented the hyperbolic orbit calculation in the Lorenz gauge with a \texttt{Mathematica} implementation that builds upon the eccentric orbit calculation from Ref.~\cite{Akcay:2013wfa}.
In the odd sector I find that Lorenz gauge source falls off as $r_p^{-3}$ and thus the code is sufficiently accurate when computing the radiated energy and angular momentum. 
For $\ell+m =\text{odd}$ I typically find that the relative error between my RWZ and Lorenz gauge implementations is $\sim10^{-8}$.
Unfortunately, in the even sector the source falls off as $r_p^{-1}$ and so it is not sufficiently accurate to compare against.
It would be necessary to derive higher-order sources, following Hopper's approach, to make it a useful benchmark.
In both the odd and even sectors though the code is sufficiently slower than the RWZ code that, at present, the Lorenz gauge code is only useful to spot check the $C^\pm_\lmo$'s for a few values of $\omega$.

Finally, as was argued in the Introduction, it is important to compare results across different methods for modelling the relativistic two-body problem.
The good agreement that I find with other methods in the next section provides further validation of my numerical SF results.

% \begin{itemize}
%     \item Show convergence of $C^\pm_{lm\omega}$
%     \item Can we envelope the convergence, which should be predictable from the source fall of, and extrapolate?
%     \item Show figure for the mode content
%     \item Discuss sources of error
%     \item Comparison with data in Table I of \cite{Faggioli:2024ugn}. The bottom value corresponds to $b=60.3216, \vinf=0.09962$.
%     \item Comparison with parabolic data in \cite{Martel:2003jj}.
%     \item Comparison with Lorenz gauge code
% \end{itemize}

% Figure from explore_hyperbolic_data_fixed_rmin_vary_vinf_NEW
\begin{figure}
    \centering
    \includegraphics[width=0.98\linewidth]{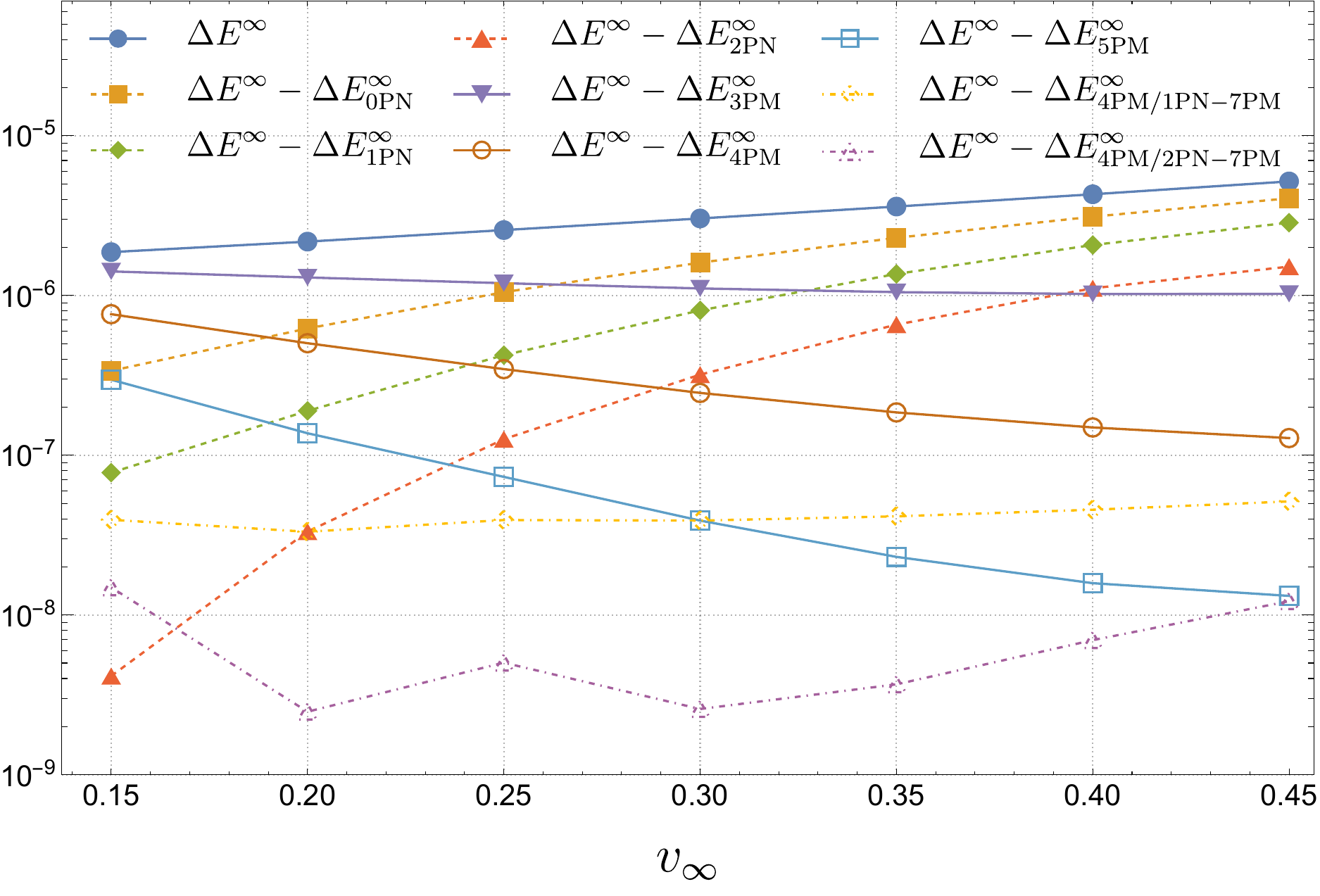}
    \caption{
		Comparison of the total radiated energy from my SF calculation with PM and PN results for $r_{\rm min} = 100m_1$.
		The PN results agree well with the SF data at low velocities but the agreement rapidly worsens for larger values of $\vinf$.
		The PM results agree well at high velocities but less well at low velocities.
		This is because for these fixed periastron orbits as $\vinf$ decreases the weak-field criterion \eqref{eq:weakfield_SF} is less well satisfied.
		The PM results improved at low velocities by adding additional PN information using the PN-PM hybrid described in Sec.~\ref{sec:hybrid}.
		I find using just 1PN information (up to 7PM) the hybrid significantly outperforms the pure 4PM result.
		By including 2PN information (up to 7PM) the 4PM hybrid is better than the 5PM result across the range of $\vinf$ plotted.
	}
    \label{fig:SF_vs_PN_vs_PM_vs_hybrid_fixed_rmin}
\end{figure}

\section{Comparison between SF, PN, PM and NR}\label{sec:results_comparisons}

% Left panel from explore_hyperbolic_data_fixed_vinf_vary_b.nb
% Right panel from explore_hyperbolic_data_fixed_vinf_vary_b_high_v.nb
\begin{figure*}
    \centering
    \includegraphics[width=0.49\linewidth]{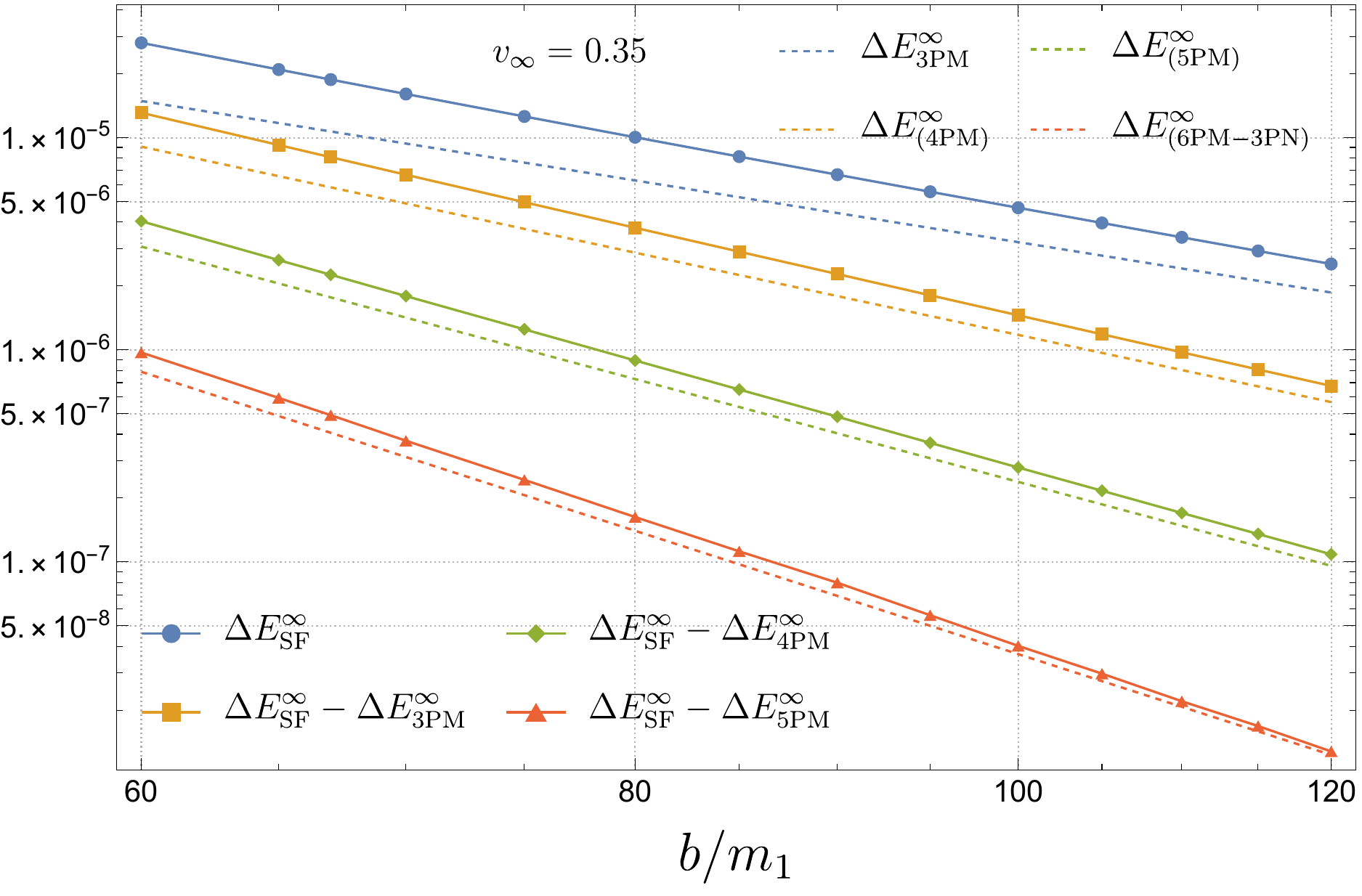}\hfill
    \includegraphics[width=0.475\linewidth]{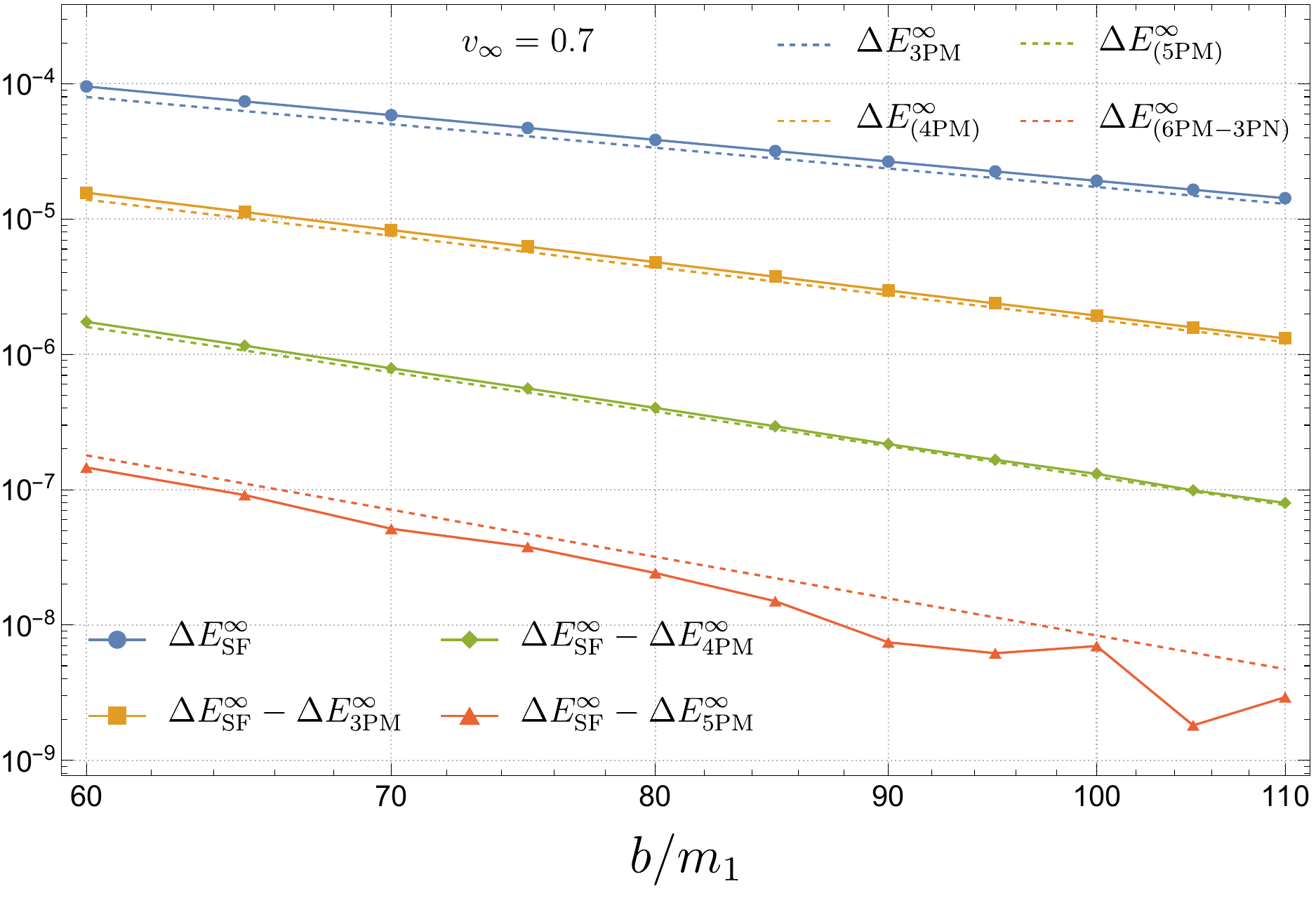}
    \caption{
		(Left panel) Comparison of the total radiated energy between SF and PM for $v_\infty/c=0.35$.
		After subtracting each PM order I find the residual has the expected scaling.
		In particular, after subtracting all highest known term (5PM) the residuals scales as $b^{-6}$.
		For a reference $b^{-6}$ curve I plot the 6PM-3PN result of Cho.
		Although the coefficient of this curve is close to the residual with 5PM, the residual with the 6PM-3PN result (not shown) does not scale as $b^{-7}$ across the range of $b$ values plotted.
		(Right panel) The same as the left panel but for $v_\infty/c=0.7$.
		There is some noise in the residual with the 5PM result but the scaling follows the $b^{-6}$ reference curve.
		This noise stems from the need to include many more $(\ell,m)$-modes for this high velocity orbit, and the high-$\ell$ modes have significantly more noise in their frequency spectrum -- see the left panel of Fig.~\ref{fig:spectrum}.
	}
    \label{fig:SF_vs_PM_fixed_vinf_vary_B}
\end{figure*}

In this section I make comparisons between my SF results with those from PN, PM, and NR.
Throughout I will plot adimensionalized quantities, e.g., unless stated otherwise $\Delta E^\infty \equiv \Delta E^\infty/ (m_1 \epsilon^2)$.
I will also often wish to distinguish between terms at a given order in a PM or PN expansion and the series \textit{through} to that order in the expansion.
For example, the 4PM contribution to the total radiated energy I will denote by $\Delta E^\infty_{(4PM)}$, whereas the expansion of the total radiated energy through 4PM (including the 3PM and 4PM terms) I will denote by $\Delta E^\infty_{4PM}$.
In this section I focus on hyperbolic orbits but I also provide results and comparisons for parabolic orbits in Appendix \ref{apdx:parabolic}.
%I will also focus on the total radiated energy, but my code computes the total radiated angular momentum as well.

First I present comparisons for the total radiated energy for orbits with a fixed periastron radius as a function of $\vinf$.
In Fig.~\ref{fig:SF_vs_PN_vs_PM_vs_hybrid_fixed_rmin} I show comparisons with PN, PM and the 4PM/1PN-7PM and 4PM/2PN-7PM hybrid models from Sec.~\ref{sec:hybrid} for an orbit with $r_{\min} = 100 m_1$.
At low velocities the PN series accurately captures the SF data, but this agreement deteriorates rapidly as the velocity of the orbit increases.
By contrast the PM series accurately captures the SF data for large velocities (the largest I consider here is $\vinf=0.45$) but becomes less accurate for low velocities.
For this fixed periastron comparison this breakdown of the PM series for low velocities occurs because the weak-field criterion \eqref{eq:weakfield_SF} is no longer satisfied at low velocity, e.g., for $r_{\min}=100m_1$ and $\vinf=0.15$ the value of $m_1/(\vinf^2b) \simeq 0.32$.
The comparison with the 4PM/1PN-7PM hybrid model in Fig.~\ref{fig:SF_vs_PN_vs_PM_vs_hybrid_fixed_rmin} improves both the low-velocity and higher-velocity results and as such significantly outperforms the 4PM results.
At low velocity the improvement stems from incorporating an approximation to the 1PN result (through 7PM order).
The comparison with the 4PM/2PN-7PN hybrid model shows further improvement and even outperforms the 5PM model across the range of velocities shown in Fig.~\ref{fig:SF_vs_PN_vs_PM_vs_hybrid_fixed_rmin}.

The results in Fig.~\ref{fig:SF_vs_PN_vs_PM_vs_hybrid_fixed_rmin} demonstrate agreement with the PM results.
I observe that after subtracting each PM order the residual is subdominant.
An even stronger test can be made if one can show that the residual scales in the expected way.
This is challenging to show in the fixed periastron radius comparison but as Eq.~\eqref{eq:DeltaE_PM} suggests fixing $\vinf$ and varying $b$ will be more illuminating.
This is because although the dependence on $\vinf$ in Eq.~\eqref{eq:DeltaE_PM} can be extremely complicated, the dependence on $b$ is simply such that after subtracting a PM result at order $k$ we expect the residual to scale as $b^{-(k+1)}$.

In Fig.~\ref{fig:SF_vs_PM_fixed_vinf_vary_B} I show the comparison for the total radiated energy between my SF data and the PM for $\vinf = 0.35$ and $\vinf = 0.7$.
In both cases, after I subtract each PM order I find the residual scales as expected.
Crucially, after I subtract the recent 5PM results, I find the residual scales as $b^{-6}$ (though I note there is some numerical noise in the 5PM residual for $\vinf=0.7$).
The 3PM and 4PM have been confirmed by multiple calculations from different groups and they also agree at low velocities with the relevant PN expansions.
Similarly, the 5PM result is found to agree with PN expansions for small $\vinf$.
To the best of my knowledge, the results presented here are the first time that the PM results have been tested at high velocity against an independent calculation.
The agreement I observe builds further confidence in the both PM and SF results.
The agreement with 5PM is especially satisfying due to the extremely complicated dependence of the 5PM coefficient on $\vinf$.

In Fig.~\ref{fig:SF_vs_PM_fixed_vinf_vary_B} the coefficient of the reference $b^{-6}$ curve comes from the 6PM-3PN result of Cho \cite{Cho:2022pqy}.
This result appears to be quite close to the numerical residual.
This is not unexpected as we see from Fig.~\ref{fig:PMvsPM3PN} that the 3PN results capture the 3PM, 4PM and 5PM results quite well up to surprising large velocities.
It is reasonable to extrapolate that a similar level of agreement would be found at 6PM order, and indeed the data supports this.
Despite the 6PM-3PN curve lying close to the 5PM residual, it is not sufficiently close to the full 6PM result that I observe a $b^{-7}$ falloff in the residual (not shown).

\begin{figure}
    \centering
    \includegraphics[width=0.98\linewidth]{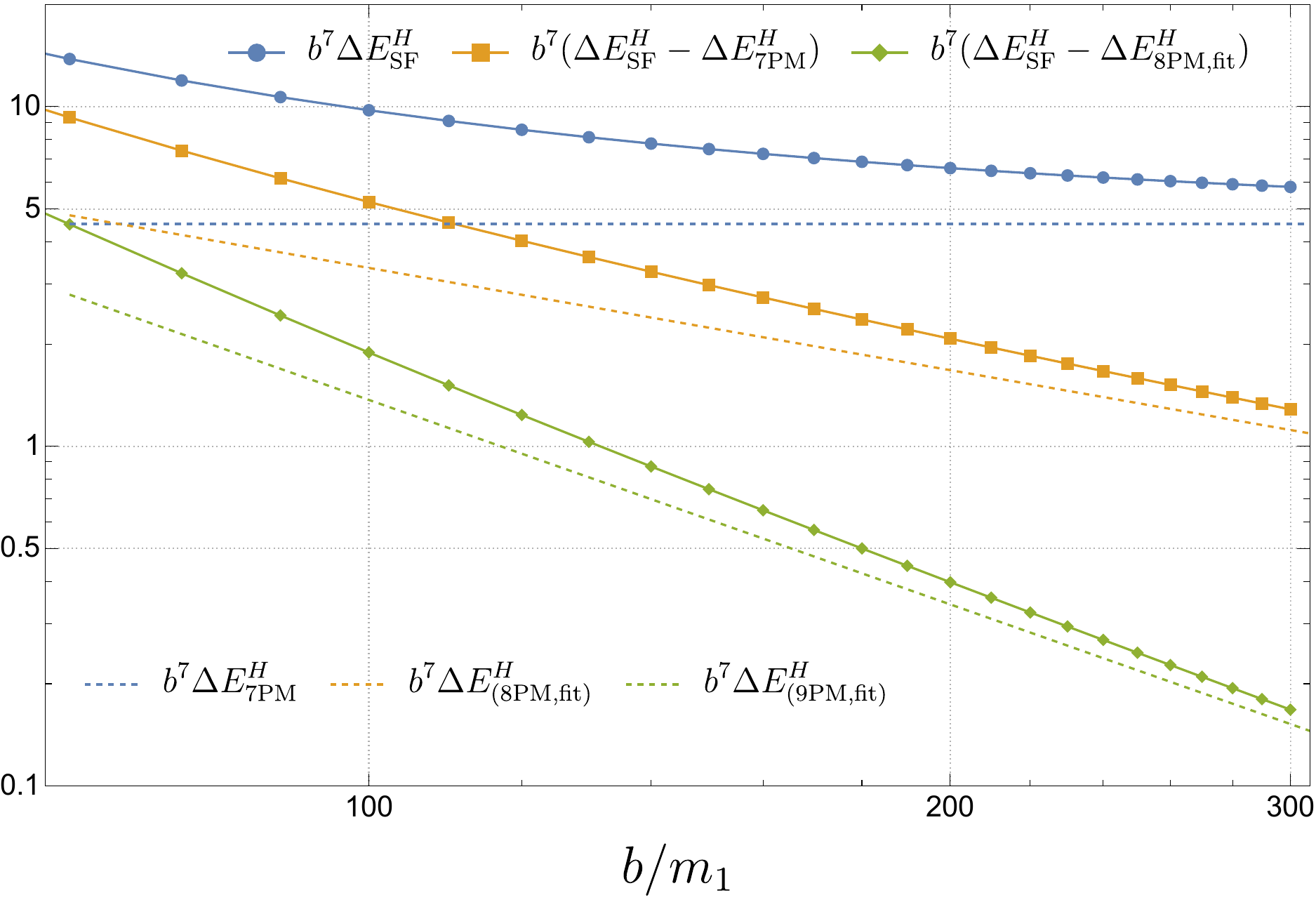}
    \caption{
		Comparison of the energy absorbed by the horizon between SF and PM for $v_\infty=0.35$.
		All the data is scaled by the $b^7$ behaviour of the leading PM result -- see Eq.~\eqref{eq:DeltaE_H_7PM}.
		After subtracting the PM result the residual scales as $b^{-8}$.
		I fit the residual as described in the main text and then further subtract the fitted $b^{-8}$ term to find a residual that scales as $b^{-9}$.
	}
    \label{fig:SF_vs_PM_horizon}
\end{figure}

Next I turn to the comparison for the horizon absorption.
Here only the leading PM term is known, which enters at 7PM order -- see Eq.~\eqref{eq:DeltaE_H_7PM}.
After subtracting this term from my SF data I find the residual scales as $b^{-8}$, as expected --- see Fig.~\ref{fig:SF_vs_PM_horizon} for an example with $\vinf=0.35$.
As I only have the leading term to compare with, the residual does not come as close the numerical noise in my data (and the numerical noise appears to be less -- compare the infinity and horizon mode spectra in the left and right panels of Fig.~\ref{fig:spectrum}, respectively).
As such, I am able to fit the residual to a function of the form $a_8 b^{-8} + a_9 b^{-9} + a_{10} b^{-10} + a_{11} b^{-11}$ where the $a_k$ are constants that are fitted for.
After further subtracting my fitted value of $a_8 b^{-8}$ I find the residual cleanly scales as $b^{-9}$.
In future work it should be straightforward to tabulate the coefficients of the 8PM and 9PM absorption up to reasonably large value of $\vinf$.

I conclude this comparison section with a first comparison of the radiated energy between SF and NR.
To the best of my knowledge, the only published results for the radiated energy from NR simulations of hyperbolic is presented in Ref.~\cite{Damour:2014afa}.
That work studies equal mass binaries with $v^*_1 = v^*_2 \simeq 0.209$. 
Using Eq.~\eqref{eq:v_and_rel_gamma}, this corresponds to a relative velocity of $v_\infty = 0.4$.
The NR simulations in Ref.~\cite{Damour:2014afa} are parameterized in terms the initial energy and angular momentum of the orbit.
Following Ref.~\cite{Long:2025tvk}, I use Eq.~\eqref{eq:b_from_E_L} to compute the corresponding impact parameter.\footnote{Note that using Eq.~\eqref{eq:b_from_E_L} results in different values of $b$ from the ones tabulated in Table I of Ref.~\cite{Damour:2014afa} as they use a different definition of the impact parameter.}
The results of the comparison between the NR and SF calculations for the radiated energy are given in Fig.~\ref{fig:SF_vs_NR}, with commentary provided in the figure caption.
In order to make the comparison I make the substitution $m_1 \epsilon^2 \rightarrow M \nu^2 +\mathcal{O}(\nu^3)$ in the formula for the self-force total radiated energy.

% \begin{itemize}
%     \item Comparison with PN and PM and NR
%     \item Extract 6PM infinity flux coefficient
%     \item Extract leading PN behaviour of the 6PM infinity flux?
%     \item Estimate the 2.5PN result?
%     \item Extract 8PM (and 9PM?) horizon flux coefficient
%     \item Include parabolic results? Yes. Here we have up to 3PN results to compare against.
% \end{itemize}

% \begin{figure}
%     \centering
%     \includegraphics[width=0.98\linewidth]{figures/DeltaE_SF_vs_PM_vinf_0.7.pdf}
%     \caption{The same as Fig.~\ref{fig:SF_vs_PM_v0.35} but for $v_\infty/c=0.7$. \NW{Here we are really plotting $\Delta E/\epsilon^2$}}
%     \label{fig:SF_vs_PM_v0.7}
% \end{figure}

% Figure from panel from explore_hyperbolic_data_fixed_vinf_vary_b.nb

\begin{figure}
    \centering
    \includegraphics[width=0.98\linewidth]{figures/Flux_hyperbolic_v_inf04_SF_PM_NR.pdf}
    \caption{
		Comparison of the radiated energy to infinity with NR and PM for $v_\infty/c=0.4$.
		The NR data comes from the equal mass binary simulations presented in Ref.~\cite{Damour:2014afa}
		The vertical, dashed line shows the location of the critical orbit in the self-force limit.
		The self-force result will diverge as $b_{\rm crit}^{\rm SF}$ is approached \cite{Barack:2026izc} but the NR data diverges at a smaller value of $b$ as the critical orbit in the NR simulations is shifted from the SF value by higher order in the mass ratio corrections.
		The NR data appears to attach quite smoothly to the SF data which suggests that, away from the critical orbit, higher order in the mass ratio corrections are small.
		The solid curves show the PM results, where I have also included a fit to the 6PM coefficient made to the SF data after subtracting the 5PM result.
		}
    \label{fig:SF_vs_NR}
\end{figure}

\section{Conclusions}\label{sec:conclusions}

In this work I made a RWZ frequency domain calculation of the gravitational radiation for a body moving along a hyperbolic orbit of a non-spinning black hole.
The main result is presented in Fig.~\ref{fig:SF_vs_PM_fixed_vinf_vary_B} where I show agreement for the energy radiated to infinity with the latest 5PM-1SF results \cite{Driesse:2024feo} in the weak field for initial velocities as high as $\vinf = 0.7$.
I also found agreement with the leading-order PM calculation of the energy absorbed by the black hole and could estimate the coefficients of the currently unknown higher-order terms -- see Fig.~\ref{fig:SF_vs_PM_horizon}.
Finally, I made comparisons with PN results, and a simple PN-PM hybrid model; see Fig.~\ref{fig:SF_vs_PN_vs_PM_vs_hybrid_fixed_rmin}, and made a first comparison with numerical relativity data in Fig.~\ref{fig:SF_vs_NR}.

The are many future directions that could build upon this work.
The code is also able to compute the radiated angular momentum, but this computation and comparison of those results with PM, PN, and NR is left for future work.
%\NW{I should update this to discuss the subtly of the calculation. Also remove the formula for $\Delta L$.}
Once the radiated angular momentum is known it would also be possible to calculate the dissipative correction to the scattering angle \cite{Damour:2020tta}.
It should also be possible to compute the nonlinear memory using the techniques of \cite{Cunningham:2024dog} and compare with recent PM results \cite{Georgoudis:2025vkk}.
The parabolic results presented in Appendix \ref{apdx:parabolic} may find utility in developing high eccentricity models for extreme mass ratio inspirals (EMRIs).
Recent work has suggested that the future LISA mission \cite{LISA:2024hlh} may detect a significant number of these sources \cite{Qunbar:2023vys,Mancieri:2024sfy}, and the latest EMRI models \cite{Chapman-Bird:2025xtd} need to be extended to cover this part of the parameter space \cite{Mancieri:2025cmx}.

% In FluxForParabolicOrbits.nb (in Cliffhanger directory)
\begin{figure}
    \centering
    \includegraphics[width=0.95\linewidth]{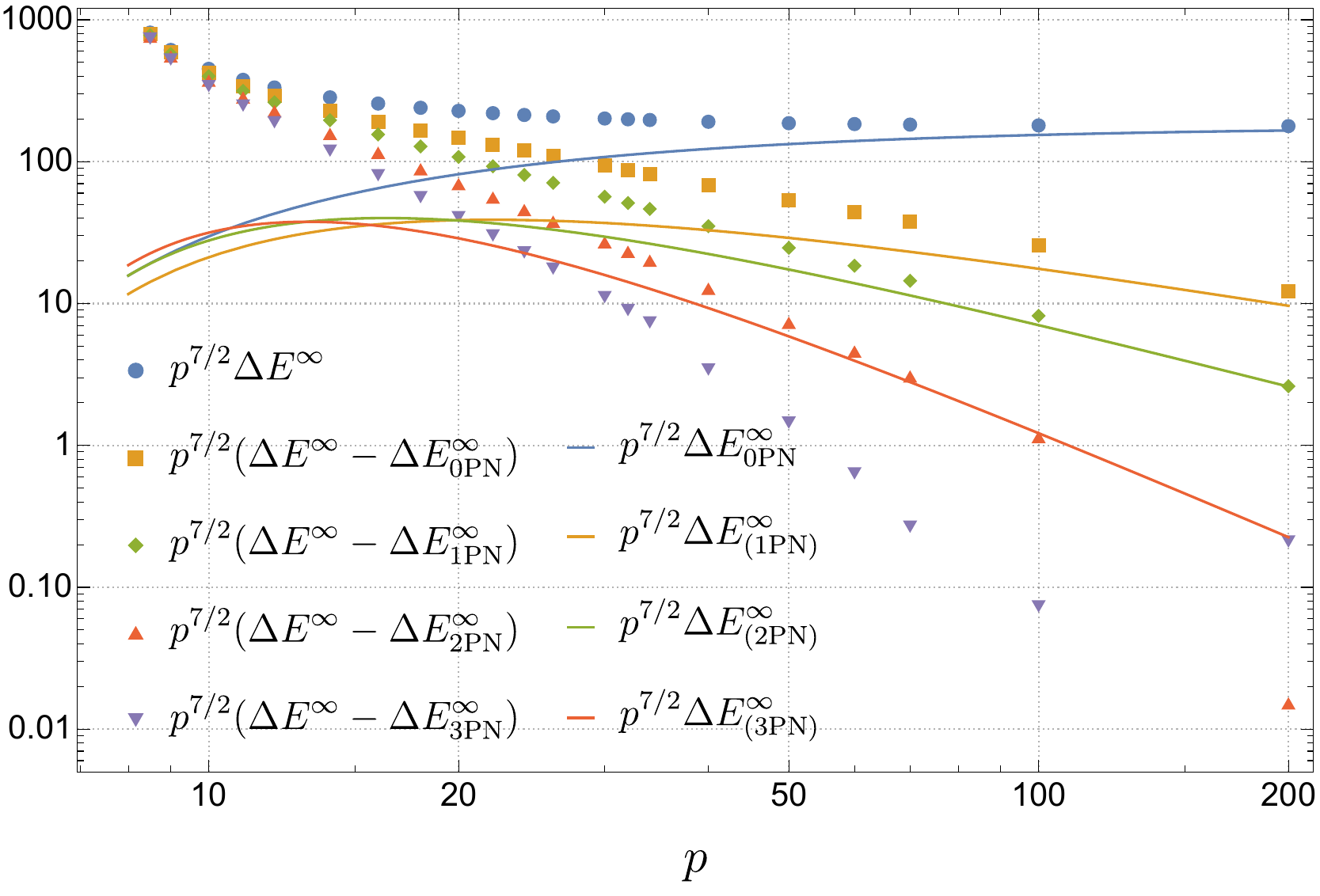}
    \caption{
	 	Results for parabolic orbits ($\E = e = 1$) and comparison with PN as a function of $p$.
		The numerical SF results are shown with (blue) circles.
		The PN results are shown with coloured curves.
		Subtracting the PN results from the coloured markers of the same colours gives residuals that follow the subdominant PN result at large values of $p$.
	}\label{fig:parabolic}
\end{figure}

There is plenty of scope for efficiency improvements of my RWZ code.
In particular, the integral to compute the asymptotic amplitudes in Eq.~\eqref{eq:weighting_coeffs} of the form where Levin's method could be applied -- see, e.g., Appendix F of \cite{Yin:2025kls} for details.
This would reduce the rapid oscillations in the source, as are apparent in Fig.~\ref{fig:cplus_convergence}, which would speed up the numerical integration.
Some of the oscillations in the source also come from the basis of homogeneous solutions that are employed when using the variations of parameters approach to computing the inhomogeneous solutions to the field equations.
It may be possible to avoid these by using spectral methods \cite{PanossoMacedo:2022fdi,Leather:2024mls}.
Yet another alternative approach would be to use the Green's function method, though so far this has primarily been developed for scalar-field perturbations \cite{Casals:2013mpa,Wardell:2014kea,OToole:2020ejc}.

This work has focused on radiative effects but the code could be used as the foundation for calculations of conservative effectives, such as the (conservative) scattering angle.
For this the local (regularized) metric perturbation along the orbital trajectory would be needed.
For a given frequency, computing this from the RWZ master functions my code calculates is straightforward \cite{Martel:2005ir, Hopper:2010uv} but the inverse Fourier transform needed to recover the time-domain solution is very poorly convergent due to the Gibbs phenomenon \cite{Barack:2008ms}.
Methods, such as extended homogeneous solutions \cite{Barack:2008ms}, have been developed to overcome this, but unfortunately they are numerically challenging for high eccentricity and hyperbolic orbits due to large numerical calculations between Fourier modes.
These issues are not insurmountable, as was demonstrated in the scalar-field case \cite{Whittall:2023xjp}, and recent work projecting Fourier modes onto an alternative basis to mitigate these issues shows promise \cite{Whittall:2025dqn}.

Finally, due to my use of the RWZ formalism, the code can only model perturbations of a non-spinning black hole.
For perturbations of spinning black holes it is natural to turn to the Teukolsky formalism \cite{Teukolsky:1973ha}.
Unfortunately, the approach developed by Hopper to accelerate the convergence of the source integral may not apply here \cite{Hopper:2017iyq}.
Instead the integration-by-parts method of Whittall and Barack \cite{Whittall:2023xjp} could be employed to achieve the same goal of increasing the rate of convergence of the source integral.
Alternatively, the Sasaki-Namura approach could be pursued and has been shown to be effective for unbound sources \cite{Silva:2023cer, Yin:2025kls}.
Calculation of either radiative or conservative SF results for orbits about a Kerr black hole would enable comparisons with the latest PM results with spinning particles, e.g., \cite{Jakobsen:2023hig}.

% Other ways to compute the fluxes:
% \begin{itemize}
%     \item Green function approach: \cite{OToole:2020ejc}
% 	\item Efficiently improvements: use Levin's method -- see Appendix F of \cite{Yin:2025kls}.
% 	\item Maybe spectral methods could avoid some of the numerical problems.
% 	\item Fit at a range of $\vinf/c$ values to estimate the 6PM coefficient. Similar for 8PM and 9PM at the horizon.
% 	\item Comparison with angular momemtnum flux and calculation of the dissipative correction to the scatting angle
% 	\item Calculation of the metric perturbation. Challenges due to poor convergence of EHS. Recent methods propose solutions, e.g., \cite{Whittall:2025dqn}
% 	\item Plan to add the code to the BHPToolkit
% 	\item Highlight how the parabolic data can help inform fits across the eccentricity parameter space (higher $e$ was called for in \cite{Mancieri:2025cmx}).
% 	\item Extension to Kerr.
% \end{itemize}

% Figure from FluxForParabolicOrbits.nb

\section*{Acknowledgements}

I thank Chris Kavanagh, Olly Long, Jakob Neef and David Trestini for helpful discussions, and Riccardo Gonzo and Sarp Akcay for comments on a draft of this work.
I also thank Hassan Khalvati for sharing data for very high eccentricity orbits.
The comparison with this data built further confidence in the parabolic orbit results in this work.
This work makes use of the Black Hole Perturbation Toolkit \cite{BHPToolkit}.
I acknowledge support from a Royal Society – Research Ireland University Research Fellowship. 
This publication has emanated from research conducted with the financial support of Research Ireland under grant number 22/RS-URF-R/3825. 

\appendix

\section{Numerical results and comparison with the literature and comparisons with PN for parabolic orbits}\label{apdx:parabolic}

In this section I present a selection of my numerical and, where available, comparison with numerical results in the literature.
Table I of Ref.~\cite{Faggioli:2024ugn} presents results for hyperbolic orbits and I make a comparison with the one weak-field result presented there.
Table III of Ref.~\cite{Martel:2003jj} presents results for parabolic orbits.
I present comparisons with these results in Table \ref{table:numerical_results}.
I also present data for hyperbolic orbits with $\vinf=\{0.35,0.7\}$ as plotted in the two panels of Fig.~\ref{fig:SF_vs_PM_fixed_vinf_vary_B}.

For parabolic orbits I also compare against the known PN series which has been computed at 1PN \cite{Blanchet:1989cu}, 2PN \cite{Bini:2020hmy} and 3PN order \cite{Cho:2022pqy}.
Note for parabolic orbits the $(b,\vinf)$ is not defined as $b \rightarrow \infty$ as the parabolic limit is approached.
Instead I parametrize the orbit with $(p,e)$, or alternatively $(\E,\calL)$.
The results of the PN-SF comparison for parabolic orbits are shown in Fig.~\ref{fig:parabolic} where the numerical data and the PN series are found to be in agreement for large values of $p$.

\begin{table*}[htb!]
    \centering
    \begin{tabular}{c c | c c | c c || >{$}c<{$} >{$}c<{$} >{$}c<{$} }
      $\E$   & $\calL$ & $b/m_1$ & $\vinf$ & $p$ & $e$ & \Delta E_\infty^{\rm here} & \Delta E_\infty^{\rm ref} & \rm{rel.~diff.} \\
      \hline
 1 & 4.08248 & - & 0 & 10 & 1 & 1.4266\times 10^{-1} & 1.4712\times 10^{-1} & 3.0\times 10^{-2} \\
 1 & 4.24264 & - & 0 & 12 & 1 & 5.5688\times 10^{-2} & 5.8467\times 10^{-2} & 4.8\times 10^{-2} \\
 1 & 4.42719 & - & 0 & 14 & 1 & 2.7657\times 10^{-2} & 2.9303\times 10^{-2} & 5.6\times 10^{-2} \\
 1 & 4.6188 & - & 0 & 16 & 1 & 1.5675\times 10^{-2} & 1.6636\times 10^{-2} & 5.8\times 10^{-2} \\
 1 & 4.8107 & - & 0 & 18 & 1 & 9.6901\times 10^{-3} & 1.0303\times 10^{-2} & 5.9\times 10^{-2} \\
 1 & 5 & - & 0 & 20 & 1 & 6.3764\times 10^{-3} & 6.7794\times 10^{-3} & 5.9\times 10^{-2} \\
 1 & 5.88348 & - & 0 & 30 & 1 & 1.3623\times 10^{-3} & 1.4374\times 10^{-3} & 5.2\times 10^{-2} \\
 1 & 6.66667 & - & 0 & 40 & 1 & 4.7204\times 10^{-4} & 4.9549\times 10^{-4} & 4.7\times 10^{-2} \\
 1 & 7.3721 & - & 0 & 50 & 1 & 2.1121\times 10^{-4} & 2.1993\times 10^{-4} & 4.0\times 10^{-2} \\
 1.005  & 6.0397 & 60.3216 & 0.0996268 & 16.937 & 1.08055 & 1.4859\times 10^{-3}	& 1.477\times 10^{-3}   & 6\times 10^{-3}  \\
 \hline
 1.06752 & 22.4179 & 60 & 0.35 & 436.464 & 7.37606 & 2.8007\times 10^{-5} & \text{-} & \text{-} \\
 1.06752 & 24.2861 & 65 & 0.35 & 513.033 & 7.98662 & 2.0952\times 10^{-5} & \text{-} & \text{-} \\
 1.06752 & 25.0334 & 67 & 0.35 & 545.376 & 8.23092 & 1.8792\times 10^{-5} & \text{-} & \text{-} \\
 1.06752 & 26.1543 & 70 & 0.35 & 595.726 & 8.59745 & 1.6070\times 10^{-5} & \text{-} & \text{-} \\
 1.06752 & 28.0224 & 75 & 0.35 & 684.543 & 9.20849 & 1.2592\times 10^{-5} & \text{-} & \text{-} \\
 1.06752 & 29.8906 & 80 & 0.35 & 779.484 & 9.8197 & 1.0046\times 10^{-5} & \text{-} & \text{-} \\
 1.06752 & 31.7588 & 85 & 0.35 & 880.55 & 10.4311 & 8.1417\times 10^{-6} & \text{-} & \text{-} \\
 1.06752 & 33.6269 & 90 & 0.35 & 987.74 & 11.0425 & 6.6897\times 10^{-6} & \text{-} & \text{-} \\
 1.06752 & 35.4951 & 95 & 0.35 & 1101.05 & 11.6541 & 5.5609\times 10^{-6} & \text{-} & \text{-} \\
 1.06752 & 37.3632 & 100 & 0.35 & 1220.49 & 12.2658 & 4.6723\times 10^{-6} & \text{-} & \text{-} \\
 1.06752 & 39.2314 & 105 & 0.35 & 1346.06 & 12.8776 & 3.9636\times 10^{-6} & \text{-} & \text{-} \\
 1.06752 & 41.0996 & 110 & 0.35 & 1477.75 & 13.4894 & 3.3908\times 10^{-6} & \text{-} & \text{-} \\
 1.06752 & 42.9677 & 115 & 0.35 & 1615.56 & 14.1013 & 2.9235\times 10^{-6} & \text{-} & \text{-} \\
 1.06752 & 44.8359 & 120 & 0.35 & 1759.5 & 14.7132 & 2.5378\times 10^{-6} & \text{-} & \text{-} \\
 \hline
 1.40028 & 58.8118 & 60 & 0.7 & 1758.02 & 29.3508 & 9.5406\times 10^{-5} & \text{-} & \text{-} \\
 1.40028 & 63.7127 & 65 & 0.7 & 2064.27 & 31.8046 & 7.3988\times 10^{-5} & \text{-} & \text{-} \\
 1.40028 & 68.6137 & 70 & 0.7 & 2395.03 & 34.2579 & 5.8522\times 10^{-5} & \text{-} & \text{-} \\
 1.40028 & 73.5147 & 75 & 0.7 & 2750.28 & 36.7107 & 4.7093\times 10^{-5} & \text{-} & \text{-} \\
 1.40028 & 78.4157 & 80 & 0.7 & 3130.03 & 39.1632 & 3.8451\times 10^{-5} & \text{-} & \text{-} \\
 1.40028 & 83.3167 & 85 & 0.7 & 3534.28 & 41.6153 & 3.1799\times 10^{-5} & \text{-} & \text{-} \\
 1.40028 & 88.2176 & 90 & 0.7 & 3963.03 & 44.0673 & 2.6596\times 10^{-5} & \text{-} & \text{-} \\
 1.40028 & 93.1186 & 95 & 0.7 & 4416.28 & 46.519 & 2.2473\times 10^{-5} & \text{-} & \text{-} \\
 1.40028 & 98.0196 & 100 & 0.7 & 4894.03 & 48.9706 & 1.9161\times 10^{-5} & \text{-} & \text{-} \\
 1.40028 & 102.921 & 105 & 0.7 & 5396.28 & 51.422 & 1.6464\times 10^{-5} & \text{-} & \text{-} \\
 1.40028 & 107.822 & 110 & 0.7 & 5923.03 & 53.8732 & 1.4254\times 10^{-5} & \text{-} & \text{-} \\
 \hline
\end{tabular}
    \caption{
		Select numerical results from my code and comparisons with results in the literature where they are available.
		The top block shows results for parabolic orbits and one strong-field hyperbolic orbit. 
		In this block the reference data for parabolic orbits is from Table III of \cite{Martel:2003jj} and reference data for hyperbolic orbits is from Table I of \cite{Faggioli:2024ugn}.
		The second and third blocks present my data for $\vinf = 0.35$ and $\vinf = 0.7$, respectively.
		This is the data plotted as the (blue) circles in Fig.~\ref{fig:SF_vs_PM_fixed_vinf_vary_B}.
	}
    \label{table:numerical_results}
\end{table*}

\bibliography{1SF-PM}% Produces the bibliography via BibTeX.

\end{document}